\documentclass[acmlarge,nonacm]{acmart}
\usepackage[utf8]{inputenc}
\usepackage{mathtools}
\usepackage{tikz}
\usetikzlibrary{arrows,chains,matrix,positioning,scopes}
\usepackage{stmaryrd}
\usepackage{listings}
\usepackage{comment}
\usepackage{subcaption}
\usepackage{xcolor}
\usepackage{algorithm2e}
\usepackage{prftree}
\usepackage{xspace}
\AtBeginDocument{%
  \providecommand\BibTeX{{%
    \normalfont B\kern-0.5em{\scshape i\kern-0.25em b}\kern-0.8em\TeX}}}

\setcopyright{rightsretained}
\acmJournal{PACMPL}
\acmYear{2021} \acmVolume{5} \acmNumber{OOPSLA} 
\acmMonth{10} \acmPrice{}\acmDOI{10.1145/3485499}

\newcommand{\bnfor}{\mid}

\newcommand{\sandor}[1]{}
\newcommand{\james}[1]{}

\newtheorem{remark}{Remark}
\theoremstyle{definition}

\newcommand{\src}[1]{\textcolor{red}{\texttt{#1}}}
\newcommand{\msrc}[1]{\textcolor{red}{#1}}
\newcommand{\id}[1]{\texttt{#1}}
\newcommand{\cor}[1]{\textcolor{blue}{#1}}
\newcommand{\mcor}[1]{\textcolor{blue}{#1}}
\newcommand{\lit}[1]{\mathit{\textcolor{black}{#1}}}
\newcommand{\sem}[1]{\textcolor{black}{\llbracket} \textcolor{red}{#1} \textcolor{black}{ \rrbracket}}
\newcommand{\sourcelang}{Pidgin\xspace}
\newcommand{\corelang}{Core Pidgin\xspace}
\newcommand{\sort}[1]{\textit{#1}}
\newcommand{\mr}[1]{\textsc{#1}}

\newcommand{\sigfun}{\mathit{sig}}

\newcommand{\KLE}{\textsc{kle}\xspace}

\title{One Down, 699 to Go: or, synthesising compositional desugarings}

\author{S{\'a}ndor Bartha}
\affiliation{%
  \institution{The University of Edinburgh}
  \streetaddress{Informatics Forum, 10 Crichton Street}
  \city{Edinburgh}
  \country{UK}}
\email{sandor.bartha@ed.ac.uk}

\author{James Cheney}
\affiliation{%
  \institution{The University of Edinburgh}
  \streetaddress{Informatics Forum, 10 Crichton Street}
  \city{Edinburgh}
  \country{UK}}
\affiliation{%
  \institution{The Alan Turing Institute}
  \streetaddress{96 Euston Rd}
  \city{London}
  \country{UK}}  
\email{jcheney@inf.ed.ac.uk}
\author{Vaishak Belle}
\affiliation{%
  \institution{The University of Edinburgh}
  \streetaddress{Informatics Forum, 10 Crichton Street}
  \city{Edinburgh}
  \country{UK}}
\affiliation{%
  \institution{The Alan Turing Institute}
  \streetaddress{96 Euston Rd}
  \city{London}
  \country{UK}}  
\email{vaishak@ed.ac.uk}
\begin{document}

\begin{abstract}
    Programming or scripting languages used in real-world systems are seldom  designed with a formal semantics in mind from the outset.  Therefore, developing well-founded analysis tools for these systems requires reverse-engineering a formal semantics as a first step. This can take months or years of effort.

    Can we (at least partially) automate this process? Though desirable, automatically reverse-engineering semantics rules from an implementation is very challenging, as found by \citet{Krishnamurthi:2019}. In this paper, we highlight that scaling methods with the size of the language is very difficult due to state space explosion, so we propose to learn semantics incrementally. We give a formalisation of Krishnamurthi et al.'s desugaring learning framework in order to clarify the assumptions necessary for an incremental learning algorithm to be feasible.
    
    We show that this reformulation allows us to extend the search space and express rules that Krishnamurthi et al. described as challenging, while still retaining feasibility. We evaluate enumerative synthesis as a baseline algorithm, and demonstrate that, with our reformulation of the problem, it is possible to learn correct  desugaring rules for the example source and core languages proposed by Krishnamurthi et al., in most cases identical to the intended rules.  In addition, with user guidance, our system was able to synthesize rules for desugaring list comprehensions and try/catch/finally constructs.
\end{abstract}




\maketitle

\section{Introduction}

Formal semantics is useful for the maintenance and analysis of programming languages, and similarly for libraries and frameworks, whose semantics might best be defined via an abstract language. But few languages are designed with formal semantics from the start, and writing formal semantics based on an already existing implementation involves guessing possible semantics rules and painstakingly writing and evaluating many different test cases. 
%Our research aims towards the semi-automation of this task.

There are many arguments in favour of reverse engineering a formal specification. Ambiguities are inevitable if the specification is only written in natural language. Various implementations may interpret the requirements differently, making it difficult to port the programs between them. Future enhancements may break properties that the creators are unaware of but that the users rely on, thus making programs unnecessarily difficult to maintain. Security risks may go unnoticed, impacting every system built with the language \cite{Amin2016}. A formal semantics is prerequisite to applying formal methods or static analysis. Moreover, the ability to reconstruct the semantics of an opaque system would make it possible to compare properties of the induced semantics with the intended design, aiding the rationalisation or redesign of the language.

The usage of formal semantics is standard in the hardware industry \cite{Kern1999}, and there is a lot of research effort to formalise aspects of mainstream languages \cite{Jung2017,Nienhuis:2016}. But its impact on the wider software industry is modest: the vast majority of systems in usage today do not have formal semantics. These systems rely on many idiosyncratic solutions to represent computations, in the form of domain specific languages, configuration languages, or query languages, and an even larger number of libraries and frameworks, that frequently change their behaviour from version to version. To get a sense of the scale of the work involved in writing the full formal semantics of one language, see some recent examples for popular programming languages, such as JavaScript~\cite{Maffeis:2008,guha10ecoop}, R~\cite{Morandat:2012}, Python~\cite{Politz:2013}, and PHP~\cite{Filaretti:2014}. Each of these formal semantics are the result of months of work by research groups.

Let us step into the shoes of the semantics engineer, whose task is to reverse engineer an interpretable semantics of an actual programming language, by observing the behaviour of its opaque or non-intelligible interpreter. Languages are usually riddled with many varieties of similar constructs, and the tedious parts of producing formal specifications involve writing a lot of small example programs, then testing their behaviour with the opaque implementation. Krishnamurthi et al., the authors of "The Next 700 Semantics: A Research Challenge" (\citeyear{Krishnamurthi:2019}) argue for the need of semi-automation. A full solution to this challenge would facilitate defining the formal semantics for 700 languages, alluding to how Peter Landin's landmark paper "The Next 700 Programming Languages" (\citeyear{Landin:1966}), written more than 50 years ago, had facilitated their creation in the first place.

As a first step towards achieving partial automation, Krishnamurthi et al. (from now on, abbreviated \KLE) suggest that semantics be divided into a complex part provided by human semantic engineers, and a tedious but shallow part hopefully synthesised automatically. They formulate this division in the following way: first let human engineers specify a core version of the language, capturing the essential features, reducing the potentially hundreds of language constructs to a few. They also provide a definitional interpreter (e.g. a direct implementation of an operational semantics) for this core language, which is much smaller. Our task then is reduced to finding translation rules from the original (source) language to this core language, based on observation of the behaviour of source programs. Note that we only work with term languages consisting of abstract syntax trees: while reverse-engineering a formal grammar for a language from its examples is also an interesting problem, we are focusing only on the semantics of the resulting term language. Our task is to find a program that translates, or \emph{desugars}, source  terms into core  terms.

For a familiar example, let the source language be a simple functional language with arithmetic. We give its semantics by first providing the semantics of the $\lambda$-calculus extended with the primitive types (numbers) and operations (arithmetic) of the language, then translating additional language constructs into this core language. Figure \ref{fig:desugar_example_tests} shows some potential test cases run on the opaque interpreter of the source language, and Figure \ref{fig:desugar_example_rule} shows an example translation rule that we should generate based on these test cases. The rule expresses a \lstinline{let} expression (in the source language) in terms of a $\lambda$ expression and application (which is part of the reduced core language). This example shows the semantics of one language construct for the sake of demonstration, but in the general case the source language may contain hundreds of constructs that need to be reduced to the core language: this reduction of the number of term constructors is the main point of the translation.

\begin{figure}[tb]
    \begin{minipage}{6.5cm}
    \centering
    \lstinline{let x=1 in x+x} \mbox{\Large $\leadsto$} \lstinline{2} \\
    \lstinline{let x=2 in 1}   \mbox{\Large $\leadsto$} \lstinline{1}%
    \subcaption{Partial input: test cases for \lstinline{let} expression}\label{fig:desugar_example_tests}
    \end{minipage}
    \begin{minipage}{7cm}
    \centering
    $\sem{\texttt{let} \, i = e_1 \, \texttt{in} \, e_2} = \Big(\mbox{\Large $\lambda$} i \, . \, \sem{e_2} \Big) \, \sem{e_1} $
    \subcaption{Partial output: translation rule for \lstinline{let} expression}\label{fig:desugar_example_rule}
    \end{minipage}
    \caption{Learning translation example}
    \label{fig:desugar_example}
\end{figure}

\KLE also suggest a natural search space: compositional \emph{desugaring} translations. They aim to synthesise a rule (or rules) for each source language term constructor. Their formulation is insightful and practical. But the specified task is still hard. They presented four unsuccessful solution attempts, describing their shortcomings and challenging the research community to overcome them. 

From their astute analysis we identify two inherent properties that make this problem so challenging: the non-standard learning framework and the intractability of the program space. The non-standard learning framework makes many search strategies hard to apply. Three out of the four attempts changed the requirements in ways the usefulness of which is questionable for an envisaged solution. Due to the astronomical search space all attempts failed to reproduce the intended semantics of their example source and core languages (which are still far behind the size of real world languages). \KLE studied a reduced search space, tree transducers. They demonstrated however with their example source and core languages that many common intended translation rules in this domain cannot be expressed by tree transducers at all.

We directly build on \KLE's work, but we take a slightly different position. We work on the principle that, before considering search strategies, the essential first step is to clearly formulate the problem to understand when a solution is even possible in principle, and then divide the task into feasible pieces. The main idea behind our paper is that the compositional translations can be learned by structuring the learning process itself compositionally (or, more precisely, incrementally). We hypothesise that we can learn the translation rules of a few term constructors at a time, in each step relying on the semantics of already established term constructors --- following human practice. This reformulation follows our intuition based on experience with semantics engineering and has the potential to help with both challenging aspects. First, it significantly reduces the search space for each synthesis sub-task, since we only synthesise a small portion of the language's translation at a time. Second, by assuming that some part of the translation is already known, we ease testing potential translations, since we can rely on the already established parts of the translation.

We argue that this partitioning into feasible sub-tasks is often possible. This paper makes the following contributions, leading to the first practical solution to \KLE's challenge:

\begin{itemize}
    \item We recapitulate  \KLE's challenge, offer further analysis of the obstacles to it, and outline our approach to overcoming them. (Section \ref{sec:research_framework})
    \item We propose a  formal framework for synthesizing compositional desugarings, %to refine the notion of compositional translations, connecting them to the notion of sublanguages. 
    define a sub-task - the \emph{desugaring extension problem}: learning the desugaring only on a portion of the source language (a sublanguage), and highlight underlying assumptions. (Section \ref{sec:rational_reconstruction})
    \item We propose a way to build search spaces for translation rules %based on representing translations 
    represented in a functional language, aiming to find a sweet spot between the too-restrictive tree transducers and the intractable search space provided by a general programming language. 
    %Starting with an even more reduced notion of translations than tree transducers, we show how to extend this search space with templates written in a general functional language. 
    We show how to express all example translation rules identified as challenging by \KLE, while retaining the ability to (relatively) efficiently enumerate programs. (Section \ref{sec:searching_for_desugarings})
    \item We test our approach by analysing a simple enumerative synthesis algorithm as a baseline, and evaluating it on case studies. In particular, we show that, using our modified specification for the task, it is possible to obtain most intended desugaring rules proposed as a test by \KLE.  We also investigated synthesizing rules for further extensions such as list comprehensions and try/catch/finally. (Sections \ref{sec:implementation} and \ref{sec:evaluation})
\end{itemize}
At a conceptual level, we analyze the desugaring synthesis problem, highlighting the importance of choices of hypothesis space; and at a technical level we provide a specific approach that can synthesize the desugarings proposed by \KLE as well as some simple extensions.

\section{Overview} \label{sec:research_framework}

\subsection{The challenge}

\begin{figure*}
\centering

\begin{minipage}{\linewidth}
\centering
\[\small\begin{array}{rcl}
    b  \in \sort{Bool} &::= &\{ \texttt{true}, \texttt{false} \}\qquad
    i  \in \sort{Id} \qquad
    n  \in \sort{Number} \qquad
    s  \in \sort{String} \\
    o  \in \sort{Op} &::= &\{ 0- , \texttt{not} ,  + ,  - , \land , \lor , < , > \} \\
    t, t_1 , t_2, t_3  \in \msrc{\sort{STerm}} &::= &
    \msrc{ \id{SFalse}  } \bnfor 
    \msrc{ \id{SNum}(\lit{n}) } \bnfor 
    \msrc{ \id{SVar}(\lit{i}) } \bnfor 
    \msrc{ \id{SStr}(\lit{s}) } \bnfor 
    \msrc{ \id{SBetween}(t_1,t_2,t_3) } \\
    &  \bnfor&
    \msrc{ \id{SPrim}(\lit{o},[t, \dots]) } \bnfor
    \msrc{ \id{SIf}(t_1,t_2,t_3) } \bnfor
    \msrc{ \id{SLam}([\lit{i},\dots ],t) } \bnfor
    \msrc{ \id{SApp}(t_1,[t_2,\dots ]) } \bnfor \\
     &  \bnfor &
   \msrc{ \id{SLet}( \lit{i} , t_1 , t_2 ) } \bnfor
   \msrc{ \id{SLetRec}(\lit{i},t_1,t_2) } \bnfor
   \msrc{ \id{SAssign}(\lit{i},t) } \bnfor
   \msrc{ \id{SList}([t,\dots ]) } \bnfor \\
   & \bnfor  &
   \msrc{ \id{SListCase}(t_1,t_2,t_3) } \bnfor 
   \msrc{ \id{SFor}(t_1,[f, \dots ],t_2) } \\
   f \in \msrc{\sort{SForBind}} &::= &\msrc{\id{SFBind}(i,t)} \\
    e, e_1 , e_2, e_3  \in \mcor{\sort{CTerm}} &::= &
    \mcor{ \id{CBool}(b)  } \bnfor 
    \mcor{ \id{CNum}(\lit{n}) } \bnfor 
    \mcor{ \id{CVar}(\lit{i}) } \bnfor 
    \mcor{ \id{CStr}(\lit{s}) } \bnfor
    \mcor{ \id{CPrim1}(\lit{o},e ) } \bnfor
    \mcor{ \id{CPrim2}(\lit{o},e_1,e_2 ) } \\
    &  \bnfor&
    \mcor{ \id{CIf}(e_1,e_2,e_3) } \bnfor
    \mcor{ \id{CLam}([\lit{i},\dots ],e) } \bnfor
    \mcor{ \id{CApp}(e_1,[e_2,\dots ]) } \bnfor
    \mcor{ \id{CLet}( \lit{i} , e_1 , e_2 ) } \\
     & \bnfor &
   \mcor{ \id{CLetRec}(\lit{i},e_1, e_2) } \bnfor
   \mcor{ \id{CAssign}(\lit{i}, e) } \bnfor
   \mcor{ \id{CList}([e,\dots ]) } \bnfor 
   \mcor{ \id{CListCase}(e_1,e_2,e_3) } 
\end{array}\]

\subcaption{Syntax of Pidgin source (\textcolor{red}{red}) and core (\textcolor{blue}{blue}) languages}
\label{fig:src_and_core}

\end{minipage}
\begin{minipage}{\linewidth}
\centering

\[\small\begin{array}{rcl}
    \sem{ \id{STrue}  } &=& \cor{\id{CBool}(true)} \\
    \sem{ \id{SFalse} } &=& \cor{\id{CBool}(false)} \\
    \sem{ \id{SNum}(\lit{n}) } &=& \cor{\id{CNum}(\lit{ n })} \\
    \sem{ \id{SVar}(\lit{i}) } &=& \cor{\id{CVar}(\lit{ i })} \\
    \sem{ \id{SStr}(\lit{s}) } &=& \cor{\id{CStr}(\lit{ s })} \\
    \sem{ \id{SBetween}(t_1,t_2,t_3) } &=& 
    \cor{\id{CLet}(\%i_1, \sem{ t_1 }, \id{CLet}(\%i_2, \sem{t_2}, \id{CLet}(\%i_3, \sem{t_3} , } \\ && \quad \quad \quad \cor{ \id{CPrim2}(\land,\id{CPrim2}(<,  \%i_1 , \%i_2 ), \id{CPrim2}(<, \%i_2  ,  \%i_3  )))))}  \\
    \sem{ \id{SPrim}(\lit{o},[t_1]) } &=& \cor{\id{CPrim1}( \lit{  o  } , \sem{  t  } ) } \\
    \sem{ \id{SPrim}(\lit{o},[t_1,t_2]) } &= &\cor{\id{CPrim2}( \lit{  o  } , \sem{  t_1  } , \sem{  t_2  } ) } \\
    \sem{ \id{SIf}(t_1,t_2,t_3) } &=& \cor{\id{CIf}( \sem{  t_1  } , \sem{  t_2  } , \sem{  t_3  } ) } \\
    \sem{ \id{SLam}([\lit{i},\dots ],t) } &=& \cor{\id{CLam}( [\lit{i},\dots ] , \sem{  t  } ) } \\
    \sem{ \id{SApp}(t_1,[t_2,\dots ]) } &=& \cor{\id{CApp}( \sem{  t_1  } , [ \sem{ t_2 },\dots ] ) } \\
    \sem{ \id{SLet}( \lit{i} , t_1 , t_2 ) } &=& \cor{\id{CLet}( \lit{  i  } , \sem{  t_1  } , \sem{  t_2  } ) } \\
    \sem{ \id{SLetRec}(\lit{i},t_1,t_2) } &= &\cor{\id{CLetRec}( \lit{  i  } , \sem{  t_1  } , \sem{  t_2  } ) } \\
    \sem{ \id{SAssign}(\lit{i},t) } &= &\cor{\id{CAssign}( \lit{  i  } , \sem{  t  } ) } \\
    \sem{ \id{SList}([t,\dots ]) } &=& \cor{\id{CList}( [ \sem{t},\dots ] ) } \\
    \sem{ \id{SListCase}(t_1,t_2,t_3) } &= &\cor{\id{CListCase}( \sem{ t_1 } , \sem{  t_2  } , \sem{  t_3  } ) } \\
    \sem{ \id{SFor}(t_1,[\id{SFBind}(\lit{i},t_3), \dots ],t_2) } &=& \cor{ \id{CApp}( \sem{ t_1 } , [ \id{CLam}([\lit{i}, \dots] , \sem{t_2}) , \id{CList}([ \sem{t_3} , \dots ]) ] ) }
\end{array}\]
\subcaption{Intended translation of the Pidgin source language into the core language}
\label{fig:translation}
\end{minipage}
\caption{Pidgin source language, core language, and desugaring}
    \label{fig:lang_example}
\end{figure*}

\KLE presented a case study highlighting the challenge of learning desugarings from examples. For convenience, we give a name to the language they proposed: \sourcelang.  Figure \ref{fig:lang_example} recapitulates the syntax of \sourcelang's source and core languages and quotes their intended translations of \sourcelang (red) to \corelang (blue). We treat the syntax of the languages (as shown by Figure \ref{fig:src_and_core}) along with their respective interpreters as the input of the learning process. The intended output is the set of translation rules that Figure \ref{fig:translation} depicts.  Except for minor notational differences, we repeat their definitions verbatim; some features, such as list "cons", are omitted but easy to add.  

\KLE did not give an explicit definition of the semantics of the core language or of the notation used for the desugaring rules. We assume the core language to be a standard call-by-value lambda-calculus with arguments evaluated left-to-right, in which all variables are assignable references.  The desugaring rules used intuitive, but not formally defined, notations for fresh names and list parameters, which we paraphrase in Figure~\ref{fig:translation}.  In the rule for $\id{SBetween}$, the notations $\%i_1, \%i_2, \%i_3$ stand for freshly generated identifiers.  In rules such as $\id{SLam}$, $\id{SApp}$, and $\id{SList}$, notations $[t,\ldots]$ and $[\sem{t},\ldots]$ stand for the (possibly empty) lists of sub-terms $t$ and their translations.  Finally, in the rule for $\id{SFor}$, the notation $[\id{SFBind}(i,t_3),\ldots]$ stands for a (possibly empty) list of bindings of identifiers $i$ to expressions $t_3$, and on the right-hand side the   notations $[i,\ldots]$ and $[\sem{t_3},\ldots]$ stand for the lists of the first and (translated) second components of the bindings, respectively (that is, the results of unzipping the list considering the bindings as pairs).   
%%%
We describe reference implementations of the core language and the desugaring rules in Appendix~\ref{appendix:reference_impl}.
%We describe reference implementations of the core language and the desugaring rules in a companion technical report~\cite{bartha21arxiv}.
%In the next section we provide a formal framework for desugaring rules that generalizes this notation.

The Pidgin source and core languages illustrate several potential complications in the modelling of translations: the presence of primitive types and operations, unrestricted number of children, and the special role of names. Handling argument lists may require expressive translation rules that can re-arrange them ($\src{SFor}$) or can pattern match on the list ($\src{SPrim}$). Fresh name generation may be needed to control the order of evaluation of arguments ($\src{SBetween})$. We will use Pidgin both for illustrating and evaluating our method. In more realistic examples the source language could be much larger than the core language, since the point of the translation is to reduce the language to a smaller one.

\subsection{Problem analysis}

\KLE gave the initial analysis of the problem and of their four solution attempts. Their first two attempts, based on tree matching and Gibbs sampling, were inspired by solutions of a similar task in natural language translation. They highlight the differences between natural language translation and this problem, then considered program synthesis techniques with their third and fourth attempts, based on genetic programming and constraint based program synthesis respectively. We pick up their thread and highlight two challenging aspects of the problem, the unusual learning framework and the vast search space, from the point of view of program synthesis.

\begin{figure}
    \centering
    \begin{tikzpicture}
  \matrix [matrix of nodes,column sep={200pt,between origins},row sep={25pt,between origins}] (s)
    { |[name=Sprog]|  \text{source program}  & |[name=Cprog]| core program  \\
      |[name=Sout]| source output  & |[name=Cout]| core output \\ };
  \draw[->,blue] (Sprog) edge node[auto] {desugar} (Cprog) ;
  \draw[->]      (Sprog) edge node[auto] {source interpreter} (Sout)  
                 (Cprog) edge node[auto] {core interpreter} (Cout) ; 
  \draw[<->,blue] (Sout) edge node[auto] {compare} (Cout) ;
\end{tikzpicture}
    \caption{Desugaring learning framework (cf.~\citet{Krishnamurthi:2019})}
    \label{fig:testing}
\end{figure}

Figure \ref{fig:testing}, quoted from their paper, captures their learning framework at a high level. The setting is  similar to an active inductive synthesis problem~\cite{jha10icse,Gulwani:2017}: we do not have a logical specification, but we can produce input/output examples from evaluations of programs by the interpreters. But there is a crucial difference: the function we are trying to learn is not the one we can directly test. In our problem statement we assumed the existence of two languages, the source and the core, and respective interpreters of the languages. We can evaluate source programs with the source interpreter and core programs with the core interpreter --- but we can only relate them to each other through the very translation we want to learn. This rules out many general program synthesis methods. 

\KLE's first two attempts assumed the availability of examples of the translation (i.e. pairs consisting of a source program and its corresponding intended translation in the core language), that are unlikely to be much easier to produce than the rules themselves. Their fourth analysed method used a constraint based program synthesis framework named SyntRec~\cite{Inala:2017}, but simplified the problem by assuming a shared output space for the two interpreters, allowing the comparison of the outputs directly, without relying on the translation that we are about to synthesise. This evades using this translation twice in the problem specification. Moreover, SyntRec also required a deep embedding of the core interpreter into the framework. This highlights that there are two slightly different problems at hand, depending on whether we treat the core interpreter as opaque (similarly to the source) or instead require that it has an implementation accessible to the synthesis algorithm. Since we assumed that the user produces formal semantics of the core language, using it in the learning process sounds reasonable. But the formal semantics of even a relatively small language is quite large for program synthesis methods to handle, and expanding this semantics in the framework results in huge constraints. This could have played a role in the explosion of the search space of the constraint based method when the languages have state. A further problem with this approach is that producing such an embedding for a synthesis method is not trivial for arbitrary formal semantics in the first place.

The second problem is shared with program synthesis in general: the astronomical size of the search space. The first two methods \KLE analysed are based on a reduced notion of compositional translations: tree transducers. They highlight with their example source and core languages that many common intended translation rules in this domain can not be expressed with tree transducers, like rules that require re-arranging lists of children ($\src{SFor}$), or fresh name generation in the \corelang expression ($\src{SBetween}$). Their third and fourth attempts, based on genetic programming and constraint based program synthesis, evade this restriction. These general program synthesis techniques, in principle, can express arbitrary computable translation rules. But applying general program synthesis methods seems even harder: the search space for arbitrary computable translations is much larger than tree transducers.

All methods failed to scale up to the full Pidgin language (which is still not the size of a real world language). We argue that this task extends the already notoriously hard program synthesis problem with a new dimension: the number of source term constructors, since, for a compositional translation, we need to synthesise a separate ``program'' for each such constructor. This results in a high number of program synthesis problems, some of which are interdependent while others can be solved independently.

The partial progress reported by \KLE highlights that the problem is very difficult. The main point of the translation is to reduce the number of term constructors, thus we need an approach that scales well along this dimension. But this could still be very difficult: not only does the search space grow exponentially by the number of term constructors, but the term constructors' translation rules may depend on each other, which makes testing translations hard. To make progress, we therefore make additional assumptions and reformulate the problem.

\subsection{Our approach}

When we teach students how to program, we typically do not tell them about all of the language features at once: this would most likely overwhelm or confuse them.  Instead, in the first lecture we start with ``hello world'' and gradually introduce related features in small groups.  Indeed, \citet{htdp} explicitly adopted this approach by providing language ``levels'', or self-contained sublanguages that intentionally exclude complex features for pedagogical purposes. 

We follow \KLE in their rational reconstruction of the problem, but we investigate a simplifying assumption. We hypothesise that the compositional translation can be learned incrementally, only learning the translation rules of a few term constructors at a time. For this strategy to work, we must assume that starting from the rules of a few term constructors given initially we can iteratively find small groups of term constructors (i.e. language ``levels'') whose translation rules can be found by only testing them on the part of the language where the translation is already established.  Our approach currently requires the user to provide this sequential decomposition of the language learning problem.
This approach yields two immediate questions. First, whether the sub-task of learning only a few term constructors is feasible. Second, whether decomposing languages into small, more easily learnable sublanguages is feasible in the first place.

We seek an answer for the first by investigating a brute force enumerative synthesis algorithm. The brute-force method of program synthesis, enumerative program synthesis, is straightforward to apply to quite general sub-tasks, so the unusual learning framework does not pose a problem. Enumerative synthesis is not necessarily slow compared to other synthesis methods: it has proven to be a very efficient technique in domains where the hypothesis space is rich and complex, but the size of programs is small (see \citet{Alur2013}). This description fits well our sub-tasks. While our enumerative algorithm is a simple brute force search, there are many candidate heuristics and pruning techniques. Enumerative synthesis combined with other methods fared well in competitions (see \citet{Alur2017} or \citet{Reynolds2019} for recent competition-winning synthesis algorithms), and may be applied to speed up our search in the future. A further consideration is that the success of enumerative synthesis directly depends on the search space. Our task is to define a search space which is expressive enough that it can express a wide variety of translation rules, including those \KLE considered challenging, but small enough for the resulting learning task to be feasible. Enumerative synthesis can demonstrate this, and could be used as both a baseline for comparison and the basis of future improvements.

To investigate the second point, we formalise our assumption, list some potential problems, and carry out a full test case: we present a solution of the Pidgin challenge problem, and two simple extensions to it, based on a (user-given) partitioning of the language.  
%We assume no additional input in the form of source to core translations. 
We also assume that the core interpreter is opaque, as discussed above. Our approach can perhaps best be characterised as a semi-automatic \emph{desugaring synthesis assistant} and does involve some additional user guidance or feedback, discussed in Section~\ref{sec:threats}.

\section{Formal framework} \label{sec:rational_reconstruction}

In this section we formalise the desugaring learning problem of \KLE and define the desugaring extension problem. We show how in principle, full desugarings can be structured as a series of extension learning problems, and investigate the conditions when such a partitioning strategy can be successful in principle.  While it is not clear how we could prove that partitioning into a sequence of desugaring extension problems is always possible, in later sections we present empirical results showing that it is possible for Pidgin.

\subsection{Syntax}

We employ multi-sorted term languages~\cite{meinke93universal} as a standard model of syntax. 
%Using a multi-sorted representation is reasonable, since the formal description of the language's abstract syntax is usually available: 
The sorts correspond to the syntactic classes in Figure~\ref{fig:src_and_core}, plus auxiliary sorts for lists and pairs. This representation allows us to limit translations to sort-preserving ones, which automatically excludes many ill-formed translation rules from the search. Our definition of the problem would work with single-sorted languages as well, but our definition of the search space relies on the sorts, and enumerative synthesis benefits greatly from restricting the search space. The usage of multi-sorted terms also highlights that in practice we may not evaluate every term, just those that belong to the sort of programs.

\begin{definition}[Signature] A signature $\Sigma$ consists of
\begin{itemize}
    \item A non-empty finite set of sorts $S_\Sigma$, with a designated sort $\sigma_p \in \Sigma$ for programs.
    \item A finite set of term constructors (function symbols) $\mathcal{F}_\Sigma$.
    \item The signature function $\sigfun : \mathcal{F} \to S^* \times S$, where $S^*$ is a (possibly empty) finite sequence of sorts. We will write $f : \sigma_1, \dots, \sigma_n \to \sigma$ (where $f \in \mathcal{F}$ and $\sigma_1, \dots, \sigma_n, \sigma \in S$) when $\sigfun(f) = (\sigma_1, \dots, \sigma_n ; \sigma)$. The number $n$ can be $0$, in which case we will write $f : \sigma$, and will call $f$ a constant symbol.
\end{itemize}

$\Sigma$ and $\Omega$ will stand for signatures $(S_\Sigma,\mathcal{F}_\Sigma,\sigfun_\Sigma)$ and $(S_\Omega,\mathcal{F}_\Omega,\sigfun_\Omega)$, respectively.
\end{definition}

\begin{remark} Note that we do not have term constructors with lists of children, therefore we need to model lists of children with an additional sort for the list and additional term constructors \lstinline{nil} and \lstinline{cons}. We do not have sort polymorphism, so we need to add a separate sort for each type of list. 

For simplicity in our model we did not include an infinite number of constants. Literals like numbers and strings therefore have their own term constructors that we left out in the simplified grammar presented in Figure  \ref{fig:src_and_core}.
\end{remark}

In our source language the sorts are \sort{Id}, \sort{Number}, \sort{String}, \sort{Op}, \msrc{\sort{STerm}}, \msrc{\sort{SForBind}}, $\texttt{List}_{\sort{Id}}$, $\texttt{List}_{\msrc{\sort{STerm}}}$ and $\texttt{List}_{\msrc{\sort{SForBind}}}$. The sorts of the core language are \sort{Bool}, \sort{Id}, \sort{Number}, \sort{String}, \sort{Op}, \mcor{\sort{CTerm}},  $\texttt{List}_{\sort{Id}}$, and $\texttt{List}_{\mcor{\sort{CTerm}}}$.  We also consider each list sort to be automatically equipped with suitable term constructors $\texttt{nil}_s : \texttt{List}_s$ and $\texttt{cons}_s: s \times \texttt{List}_s \to \texttt{List}_s$.  We consider that sort $\sort{SForBind}$ is essentially a pair of an identifier and $\sort{STerm}$, which we may also write as $\sort{Id} \times \sort{STerm}$.

\begin{definition}[Abstract language] An \emph{abstract language} (or language for short) over signature  $\Sigma$ is the set of terms defined inductively with the term constructors in the signature. We write $T_{\Sigma}$ for this set. The \emph{size} of a term is the number of term constructors in it. We extend the signature function $\sigfun$ to terms, we will use the same notation for the extended function, and we will use $T_{\Sigma}^\sigma$ for terms that belong to the sort $\sigma$.
We  call $T_\Sigma^{\sigma_p}$ the set of program terms.
\end{definition}

In functional programming, terms of a language can be represented with an algebraic datatype (defining a separate type for each sort in the signature). In universal algebra terminology, an abstract language is the free multi-sorted algebra over the signature $\Sigma$.

\subsection{Semantics}

As explained earlier, we model both the source and core language as black-box interpreters, which might seem natural to model as mathematical functions from terms to  result values. However, this approach is too simplistic, since there are several complications that may arise:

\begin{itemize}
    \item The interpreter may not be deterministic.
    \item The output space may not be part of the (input) language: it may contain opaque functions, locations, or other non-observable values.
    \item The observable behaviour may not be reduced to an input-output pair: there could be other side effects like I/O operations. 
    \item Similarly, we may get errors because the source language is not defined on all terms.  Some terms may be allowed by the sort system, but still result in some kind of error. For example, in our source language the grammar permits nonsensical expressions like $\src{\id{SPrim}}(<,[])$.
    \item Some computations may not terminate. We may model the output of diverging computations with $\bot$, but for Turing-complete languages, the evaluation function is not computable.
\end{itemize}

A standard way in semantics and pure functional programming to model these complications is using \emph{monads}~\cite{Moggi91,Wadler92}. To make sense of the composition of operations in the learning setting (see Figure \ref{fig:testing}), we may assume that  the source interpreter, the core interpreter, and the translation we want to learn all live in the same monad $\mathcal{M}$. %But we do not know how to extend learning to arbitrary monads, and start from a much simplified setting. 
We only consider deterministic languages, and we assume that output values (of a successfully terminating program) are part of the input language. In the model we assume that the evaluation function is computable, and terms where evaluation exhausts resources (which includes time or CPU cycles and memory) evaluate to a specific error ($\bot$).  We assume for simplicity that for a given interpreter there is a fixed (and sufficiently large) resource bound such that the interpreter can be modelled as a deterministic, computable function.   %This makes the function computable, but it may result in non-determinism, since running time may not be deterministic, and replacing the output with $\bot$ when the computation takes too much time reflects the running time in the observable output. We ignore this, and still assume that the evaluation function is deterministic.
Moreover, we assume that side-effects on any state maintained by $\mathcal{M}$ (e.g. variable assignments) are not directly observable; we only observe the final output value.

What we get through these simplifications is an interpreter that maps terms to either values (i.e. fully-evaluated terms) or errors. We assume that the set of errors $E$ is shared between the source and core languages to allow comparing the outputs. In other words, our computations take place in the error monad over the category of total computable functions. We give a simplified definition that suffices for our purposes:

\begin{definition}[Error monad] Let $E$ be a fixed, finite set of errors that contains $\bot$ and any errors the interpreter may return (that are not part of the input language), like \lstinline{syntax_error}. We assume that $E$ is disjoint from any set of terms in any languages, and let $\uplus$ mean the disjoint union of two disjoint sets. Our interpreters and translations (and translation rules) will be functions of the form
 $f : A \to B \uplus E $.
We abuse notation by defining a notion of composition for such functions by propagating the error as follows. Let $f : A \to B \uplus E$ and $g : B \to C \uplus E$. Then
$$ \forall a \in A, g(f(a)) = 
    \begin{cases*}
      g(b) & if $f(a) = b \in B$ \\
      e    & if $f(a) = e \in E$
    \end{cases*}
$$

In the terminology of category theory this amounts to saying that these functions are morphisms of the Kleisli category of the error monad and are composed as such.

We also silently propagate the error from the left when a function has multiple arguments. Let $f_1 : A_1 \to B_1 \uplus E$, $f_2 : A_2 \to B_2 \uplus E$ and $g : B_1 \times B_2 \to C \uplus E$. Then
$$ \forall a_1 \in A_1, a_2 \in A_2, g(f_1(a_1),f_2(a_2)) = 
  \begin{cases*}
  e_1 & if $f_1(a_1) = e_1 \in E$ \\
  e_2 & if $f_1(a_1) = b_1 \in B_1$ but $f_2(a_2) = e_2 \in E$ \\
  g(b_1,b_2) & if  $f_1(a_1) = b_1 \in B_1$ and  $f_2(a_2) = b_2 \in B_2$
  \end{cases*}
$$

We extend this shorthand to handle an arbitrary number of arguments.

\end{definition}

\begin{definition}[Semantics] 

An \emph{evaluation function} (or \emph{interpreter}) of a language $T_{\Sigma}$ is a function $\phi : T_{\Sigma}^{\sigma^p} \to T_{\Sigma}^{\sigma^p} \uplus E$ over the terms such that:

\begin{itemize}
    \item $\forall t \in T_{\Sigma}^p, \phi(\phi(t)) = \phi(t)$, that is, $\phi$ is idempotent.
\end{itemize}

The members of the image of $\phi$ in $T_{\Sigma}^p$ (that is, the image of $\phi$ apart from the errors $E$) are called the \emph{values} of the language, denoted by $V(T_{\Sigma})$.  A value evaluates to itself (cf. Remark~\ref{rem:values}).

\end{definition}

\subsection{Translations}

%We hope that the formal semantics of the source language can be defined by a translation to the core language, and we want to automatically synthesise such a definition. 

In general, the sorts of the source and the core language are different, and indeed, in our example the core language has Booleans but the source does not, while the source language has a sort for $\msrc{\sort{SForBind}}$ and the added $\texttt{List}_{\msrc{\sort{SForBind}}}$, and neither has a corresponding sort in the core language. Our methods to build a search space require a partial mapping from the source sorts to the core sorts, that preserves the sort of programs. 

%We simplify the situation by assuming that the set of sorts and the sort of programs are the same for the source and core languages. This simplification covers the case when the partial mapping from the source sorts to the core sorts is injective: in this case we can identify the sorts paired by the mapping and add sorts without a pair into the other language as empty sorts without any change of the languages (the set of terms). This is the case in our example languages: we identify \msrc{\sort{STerm}} with \mcor{\sort{CTerm}} and their corresponding lists as the same sorts (but with different term constructors in the respective languages).

%The more general case when the mapping between sorts is not injective is subtle. In principle, our search methods would still work, but could be less efficient, since we lose information. The sorts partition the terms of the language, and when we search for a sort-preserving translation of a source term we only need to consider terms in the core language that belong to the corresponding sort: the less fine grained this partitioning the larger will be the search space. This generalisation seems unnecessary for our current goals and is beyond the scope of the paper. 

We assume for simplicity that the sorts of the source and core languages are subsets of a common set $S$, and that the sort of programs is common to the two languages.
Another case that we may consider is when the core language has an expressive type system, and all source language sorts are mapped to some type in the core language, in other words the source language is embedded into a typed core language. We also leave this variant of the problem for future investigations.

Translations take place in the same error monad as interpreters: this lets us compose them with the interpreters. Translations can naturally return errors. If evaluating a source term results in a syntax error, then we may not want to map it into the core language, we may simply want to translate it directly into the syntax error. For example, our intended translation in Figure \ref{fig:src_and_core} does not give a mapping for a term with constructor $\src{SPrim}$ and zero arguments. We may imagine the source interpreter raising a syntax error in this case, and it is more natural for the translation to raise the same syntax error directly, since in the core language there are no corresponding constructs that result in a syntax error in the core interpreter. 

\begin{definition}[Translation] Let us have %two languages, 
a source language $T_\Sigma$ and a core language $T_\Omega$, where the signatures share the set of sorts $S$. We will refer to sort-preserving functions $\delta : T_\Sigma \to T_\Omega \uplus E$ as \emph{translations}.
\end{definition}

\begin{definition}[Interpretation] Let $T_\Sigma$ be the source and $T_\Omega$ the core language, again with a common set of sorts. Let $f \in \mathcal{F}_\Sigma$ be a term constructor with signature  $f : \sigma_1 \times \dots \times \sigma_n \to \sigma$. Then we will call members of the function space
$T_\Omega^{\sigma_1} \times \dots \times T_\Omega^{\sigma_n} \to T_\Omega^{\sigma} \uplus E$ \emph{interpretations} of $f$ over the signature $\Omega$. When $f : \sigma$ ($f$ is constant), then an interpretation is simply an element of $T_\Omega^{\sigma} \uplus E$.
\end{definition}
%fst(Just(x,y)) = x
%fst(Nothing) = Nothing
%etc.
%
%pair :: (A + E -> B + E) -> (A + E -> C + E) -> (A + E -> (B,C)+E)
%pair f g = \x. 
%  case x of Nothing -> Nothing 
%            Just(a) -> case (f(a), g(a)) of (Just(b),Just(c)) -> Just(b,c) 
%                                            _ -> Nothing
%
%

\begin{definition}[Compositional translation] Let $T_\Sigma$ be the source language and $T_\Omega$ be the core language. An \emph{interpretation} of the source signature $\Sigma$ into the target language $T_\Omega$ is a function $\pi$ that assigns an interpretation to every term constructor $f \in \mathcal{F}_\Sigma$ over the signature $\Omega$.  We refer to $\pi(f)$ as a \emph{translation rule} for $f$ in $\pi$.

An interpretation of the source signature defines a sort-preserving function $\delta : T_\Sigma \to T_\Omega \uplus E$ from source to core terms using structural recursion. We will call such translations  \emph{compositional}. In functional programming terms these are ``folds'' (or more precisely, traversals), over the datatype defined by $\Sigma$.
%, and in universal algebra terms these are just multi-sorted $\Sigma$-algebras with carrier $T_\Omega$.
\end{definition}

\begin{remark} Note that in our intended translation of the example languages \src{\id{SPrim}} has two rules. This poses no problem in \KLE's original model: tree transducers, if the tree transducer is not deterministic. But in our model we always assume one translation rule per term constructor, and the rule for \src{\id{SPrim}} can only be expressed by case analysis on lists.
\end{remark}

\subsection{Correctness}

\begin{definition}[Soundness]
Let $T_\Sigma$ and $T_\Omega$ be two languages, with two evaluation functions $\Phi_\Sigma$ and $\Phi_\Omega$, respectively. A translation $\delta : T_\Sigma \to T_\Omega \uplus E$ is \emph{sound} iff (cf. Figure~\ref{fig:testing})
$$\forall t \in T_\Sigma^p, \Phi_\Omega(\delta(t)) = \delta(\Phi_\Sigma(t))$$
\end{definition} 

\begin{remark}\label{rem:values}
A sound translation maps values to values or errors: if $v \in V(T_\Sigma)$, then $\Phi_\Sigma(v) = v$, and since $\delta$ is sound, $\Phi_\Omega(\delta(v)) = \delta(\Phi_\Sigma(v)) = \delta(v)$, which means that $\delta(v)$ is either a value or an error.
\end{remark}

\begin{definition}[Adequacy]
Let $T_\Sigma$ and $T_\Omega$ be two languages, with two evaluation functions $\Phi_\Sigma$ and $\Phi_\Omega$, respectively. A translation $\delta : T_\Sigma \to T_\Omega \uplus E$ is \emph{adequate} iff
$$\forall v \in V(T_\Sigma) ,  \delta(v) \in V(T_\Omega) \qquad \text{and} \qquad
\forall v_1, v_2 \in V(T_\Sigma) , v_1 \ne v_2 \implies \delta(v_1) \neq \delta(v_2)$$
that is, $\delta$ maps values to values and it is injective when restricted to values.
\end{definition}

\begin{definition}[Correctness]
We will call a sound and adequate translation \emph{correct}.
\end{definition}

In practice we cannot test these conditions, provided the source language is Turing-complete, by Rice's theorem since correctness is  a nontrivial program property. We define notions approximating correctness with a set of source programs: a \emph{test set}. In a more general setting, we may assume that --- putting non-termination aside --- we can choose to evaluate arbitrary source language terms during the algorithm. Our approach abstracts away from the choice of the test set.

\begin{definition}[Correctness with respect to a test set] Let $T_\Sigma$ and $T_\Omega$ be two languages with evaluation functions $\Phi_\Sigma$ and $\Phi_\Omega$, respectively. Let $\mathcal{I} \subset T_\Sigma^p$ be a subset of programs of the source language. A translation $\delta : T_\Sigma \to T_\Omega$ is \emph{sound with respect to the test set} $\mathcal{I}$ iff
$$\forall t \in \mathcal{I}, \Phi_\Omega(\delta(t)) = \delta(\Phi_\Sigma(t))$$

A translation  is \emph{adequate with respect to the test set} $\mathcal{I}$ iff
$$\forall t \in \mathcal{I}, \Phi_\Sigma(t) \notin E \implies \delta(\Phi_\Sigma(t)) \notin  E \qquad \text{and}$$
$$\forall t_1, t_2 \in \mathcal{I}, \Phi_\Sigma(t_1) \neq \Phi_\Sigma(t_2) \implies \Phi_\Omega(\delta(t_1)) \neq \Phi_\Omega(\delta(t_2))$$

A translation is \emph{correct with respect to the test set} $\mathcal{I}$ iff it is sound and adequate with respect to $\mathcal{I}$.
\end{definition}

\begin{remark}
Although we assumed above that all values are representable, languages with non-representable values could be handled by requiring that the example programs do return a representable value.  A more difficult question is how to deal with nondeterminism or observable side-effects (e.g. due to concurrency or I/O); as far as we know all approaches to tested semantics rely on repeatable (i.e. deterministic) tests.
\end{remark}
\subsection{Sublanguages}

Our main idea is that compositional translations can naturally be partitioned along with the language. First we define the notion of sublanguage and language extensions, and then the extension of compositional translations.

\begin{definition}(Sublanguage) A signature $\Sigma'= (S,\mathcal{F}_{\Sigma'},\sigfun_{\Sigma'})$ is a \emph{subsignature} of signature $\Sigma = (S,\mathcal{F}_\Sigma,\sigfun_{\Sigma})$, denoted as $\Sigma' \subseteq \Sigma$, if 
$$\mathcal{F}_{\Sigma'} \subseteq \mathcal{F}_\Sigma 
\qquad \text{and} \qquad
\forall f \in \mathcal{F}_{\Sigma'}, \sigfun_{\Sigma'}(f) = \sigfun_\Sigma(f)\,.$$

Let $\Phi_\Sigma$ be the evaluation function of the language $T_\Sigma$. We will call $T_{\Sigma'}$ the \emph{sublanguage} of $T_\Sigma$, if it is closed with respect to the evaluation function, that is:
$$\Sigma' \subseteq \Sigma \qquad \text{and} \qquad \forall t \in T_{\Sigma'}, \Phi_\Sigma(t) \in T_{\Sigma'} \uplus E \,.
$$
We will also call $\Sigma$ an \emph{extension} of $\Sigma'$ and similarly $T_\Sigma$ an extension of $T_{\Sigma'}$.

\end{definition}

\begin{definition}[Extension of a compositional translation] Let $T_{\Sigma'}$ be a sublanguage of $T_\Sigma$, and let $T_\Omega$ be our target language. Let $\pi'$ be an interpretation of $\Sigma'$ into $T_\Omega$. We will call an interpretation $\pi$ of $\Sigma$ into $T_\Omega$ an \emph{extension} of $\pi'$ if it assigns the same interpretation to every term constructor in the sub-signature:
$$\forall f \in \mathcal{F}_{\Sigma'}, \pi'(f) = \pi(f)$$

We will also call the compositional translation $\delta$ defined by $\pi$ an \emph{extension} of $\delta'$ defined by $\pi'$.
\end{definition}

We intend to consider different \emph{hypothesis spaces} (i.e. sets of possible desugaring rules to consider). We also might want to consider different hypothesis spaces for different language extensions.

\begin{definition}[Hypothesis space] Let us fix $T_\Sigma$ as our source language and  $T_\Omega$ as our target language. Let $T_{\Sigma'}$ be a sublanguage of $T_\Sigma$. The hypothesis space corresponding to the signatures  $\Sigma$,  $\Sigma'$  and $\Omega$ is an enumerable set of interpretations (into $T_\Omega$) for each term constructor $f \in \mathcal{F}_\Sigma \setminus \mathcal{F}_\Sigma'$.

We will use $\mathcal{H}$ to stand for our chosen hypothesis space, and $\mathcal{H}^f$ for the set of interpretations the hypothesis space assigns to the term constructor $f$.

When the source language, its sublanguage, the target language, the interpretation of the sublanguage into the target language and a hypothesis space is understood, we will call a compositional translation that is an extension of the interpretation of the sublanguage and generated by interpretations belonging to the hypothesis space a \emph{desugaring}.

\end{definition}

\subsection{The desugaring extension problem}

We can finally define the sub-task: extending a desugaring from a sublanguage to a larger portion of the language.

\begin{definition}[Desugaring extension problem] The learning task is defined as follows:
\paragraph*{Inputs:} 
\begin{itemize}
    \item A signature $\Sigma$ of the source language, and a signature $\Omega$ of the target language.
    \item A finite test set of example input terms of the source language: $\mathcal{I} \subset T_\Sigma^{\sigma_p}$, and their corresponding outputs according to an evaluation function $\Phi_\Sigma$.
    \item A black-box evaluation function $\Phi_\Omega$ for the language $T_\Omega$.
    \item A subset signature $\Sigma' \subset \Sigma$, that defines a sublanguage $T_{\Sigma'}$
    \item A hypothesis space $\mathcal{H}$ for the sublanguage $T_\Sigma'$.
    \item A correct translation defined on the sublanguage: $\delta' : T_{\Sigma'} \to T_\Omega$.
\end{itemize}

\paragraph*{Output:} A desugaring $\delta : T_\Sigma \to T_\Omega$, such that:
 \begin{itemize}
     \item $\delta$ is an extension of $\delta'$
     \item $\forall f \in \mathcal{F}_\Sigma \setminus \mathcal{F}_{\Sigma'}$, the translation rule of $f$ in $\delta$ belongs to the hypothesis space $\mathcal{H}^f$
     \item $\delta$ is correct with respect to the test set $\mathcal{I}$. 
 \end{itemize}

\end{definition}

In the desugaring extension problem we have the interpretation of the sublanguage's term constructors as input, and we are searching for the rest of the term constructors' interpretations. The main differences between our task and the task described (in an informal manner) by \KLE are:

\begin{itemize}
    \item We assume that the desugaring rules may be partially known.
    \item We explicitly use multi-sorted terms and sort-preserving translations.
    \item We assume that the test set of source programs $\mathcal{I}$ is given as input, and the source language interpreter may not be called on inputs outside of $\mathcal{I}$.
    %\item We do not allow invoking the source interpreter during the search.
\end{itemize}

The first two points restrict the search space so that we can define feasible tasks, and the last one simplifies the setting of our search problem.

Our example task as depicted in Figure \ref{fig:lang_example} can also be seen as an instance of the desugaring extension problem. The sublanguage $T_{\Sigma'}$ is the language only containing literals (numbers, strings), identifiers and operation symbols, as their translation is fixed in advance (not included in the intended translation); the extended language is the full Pidgin source language. However, to the best of our knowledge it is not known how to solve this problem as the search space is too large. Therefore we divide it into a series of desugaring extension problems.

\subsection{Sequential learning}

Let us assume that we are looking for the full desugaring of a source language $T_\Sigma$ into a target language $T_\Omega$. The user should divide the language into an increasing series of sublanguages:
$$\Sigma_0 \subset \Sigma_1 \subset \dots \subset \Sigma_n = \Sigma$$
where, for every $n \in [1\dots n]$ , $T_{\Sigma_{n-1}}$ is a sublanguage of $T_{\Sigma_n}$. Assume that we know the translation for $\Sigma_0$ (which typically contains literals and operations for primitive types). The user also needs to provide a hypothesis space $\mathcal{H}_i$ and a test set $\mathcal{I}_i$ for each sub-task.

Right now we do not investigate how we can obtain suitable partitioning of the language, or suitable test sets; we assume they are part of the user input.

\begin{definition}[Sequential desugaring learning problem] The definition of the full learning task:
\paragraph*{Inputs:}\begin{itemize}
    \item A signature $\Sigma$ of the source language, and a signature $\Omega$ of the target language.
    \item A series of sub-signatures: $\Sigma_0 \subset \Sigma_1 \subset \dots \subset \Sigma_n = \Sigma$ where, for every $k \in [1\dots n]$ , $T_{\Sigma_{k-1}}$ is a sublanguage of $T_{\Sigma_k}$.
    \item A series of finite test sets $\mathcal{I}_k \subset T_{\Sigma_k}$ for every  $k \in [1\dots n]$, and their corresponding outputs according to an evaluation function $\Phi_\Sigma$.
    \item A black-box evaluation function $\Phi_\Omega$ for the core language $T_\Omega$.
    \item Hypothesis spaces $\mathcal{H}_1,\ldots,\mathcal{H}_n$.
    \item A desugaring defined on the minimal sublanguage: $\delta_0 : T_{\Sigma_0} \to T_\Omega$ that is correct on $T_{\Sigma_0}$.
\end{itemize}

\paragraph*{Output:} A desugaring $\delta : T_\Sigma \to T_\Omega$, such that 
\begin{itemize}
    \item $\delta$ is an extension of $\delta_0$
    \item $\forall k \in [1\dots n], \forall f \in \mathcal{F}_{\Sigma_k} \setminus \mathcal{F}_{\Sigma_{k-1}}$, the translation rule of $f$ in $\delta$ belongs to the hypothesis space~$\mathcal{H}^f_k$
    \item $\delta$ is correct with respect to the full test set $\bigcup_{k \in [1\dots n]} \mathcal{I}_k $.
\end{itemize}

\end{definition}

Note that the sequential desugaring learning problem is almost the same as the desugaring extension problem, we merely added some additional inputs that divide the search space, so we partition one desugaring extension problem into a series.  

\subsection{Conditions of a solution}

We conclude this section with a short discussion of research questions entailed by our approach.

%We finish the rational reconstruction with a short remark that our tasks in general may not have a solution, and if a solution exists it may not be unique.

\paragraph{Do solutions exist?  Can we solve sequential problems by composing solutions one extension at a time?}

Some desugaring extension problems may not have a solution, since the chosen hypothesis space may not contain the intended desugaring. But even if a correct desugaring exists, it is possible that the given partial desugaring (while being correct on the sublanguage) can not be extended to a correct full desugaring.
%Assume that a correct desugaring exists, that is, there is a translation in the hypothesis space which translates the source language correctly into the target language. This still does not guarantee the existence of a solution, because it is possible that the given partial desugaring can not be extended to a correct full desugaring, regardless that the partial desugaring is correct on the sublanguage. 
The reason is that some state or value may not be accessible in the sublanguage, which makes some globally bad translations correct when only tested on the sublanguage. For example, desugaring a conditional \texttt{if X then Y else Z} expression into \texttt{Y} could be correct in the sublanguage, if in the sublanguage all Boolean expressions evaluate to true. Further examples are desugaring a try-catch expression simply to the main body in a sublanguage without exceptions, or desugaring a $\texttt{fst}$ selector of a pair to an error value in a sublanguage without pairs.  This means that a naive, greedy strategy for solving a sequential problem might not work: we might commit to the wrong semantics for a sublanguage too early, making later extensions impossible. These silly examples seem pathological,  but in practice, it seems likely that a work-in-progress semantics will be in such a ``stuck'' state most of the time, so understanding and finding ways of mitigating this situation is a major challenge.  

\paragraph{Are solutions unique?}

In program synthesis, especially in inductive (example-based) synthesis, it is a common problem that there are multiple programs that satisfy the specification (see \cite{Gulwani:2017}, chapter 7.4). Our task is similar: not only could semantically different translations be correct with respect to the finite test set, but even correct translation rules often can be expressed multiple, semantically equivalent ways. These equivalent programs can often be transformed into each other by semantics-preserving transformations, like swapping the two branches of a conditional expression and negating the condition, etc. Our task has a special ambiguity of this kind, that not only concerns the individual translation rules but the whole translation. The reason is that languages often have (many) \emph{symmetries}: that is, correct, invertible translations into themselves.  A composition of two correct translations is still correct, so such symmetries induce many equivalent but syntactically different desugarings. 
Swapping two branches of the conditional is just the special case of the well-known symmetry of the Boolean language that exchanges true and false. It is possible to map Boolean constants to the opposite values, switch the \texttt{and} and \texttt{or} logical operations, and switch the order of the branches of the \texttt{if} conditional, extending the transformation form an individual translation rule to the whole translation.

%One example is the well-known symmetry of the Boolean language that exchanges true and false. It is possible to map Boolean constants to the opposite values, switch the \texttt{and} and \texttt{or} logical operations, and switch the order of the branches of the \texttt{if} conditional. These transformations not only can transform individual translation rules into equivalent ones, but a whole translation can be transformed in which the source \texttt{true} constant is mapped to the core \texttt{false} constant, etc.

\paragraph{Can we devise test sets that ensure correctness?}

It is also possible that there are multiple, semantically inequivalent solutions that are correct with respect to the given test set, but not all of them are correct with respect to the whole source language. A question that arises is whether there a finite test set $\mathcal{I}$ such that if a translation is correct with respect to $\mathcal{I}$ then it is correct with respect to the whole source language? Assuming opaque evaluation functions such a test set may not exist.  What reasonable assumptions can we make about the languages to ensure that such test sets exist? Moreover, when do small test sets exist (where the size of a test set is the sum of the size of its elements), so that we can find a correct desugaring not just in theory, but in practice with extensive testing?  In particular, since our approach currently places the burden of designing a good test set on the user, it is an interesting question whether (and how) we could also automate the process of synthesising such test sets, perhaps in a counterexample-guided inductive synthesis (CEGIS) loop~\cite{solar-lezama05pldi,jha10icse}.

\paragraph{Can desugaring learning problems be decomposed into feasible sequential extension problems?}

A final, and perhaps most important, question is: what is the exact semantic condition for the existence of a partitioning of the source language into a series of sublanguages such that each desugaring extension problem is feasible and collectively they lead to a solution of the full problem?  For example, such that each extension is small (contains less than $n$ term constructors for some small $n$), extensions of partially correct translations exist in each step, and the extension can be found with a small test set.  This seems like an inherently empirical and language-dependent question and in the rest of this paper we present experimental results investigating this question for the challenge problem.

\section{Searching for desugarings} \label{sec:searching_for_desugarings}

\subsection{The algorithm}

We use a simple enumerative synthesis algorithm to solve our sub-tasks (the desugaring extension problems).  A sequential problem is solved greedily, using the solution in each step as the input desugaring to the next step.

% \begin{algorithm}[tb]
% \small
% \KwIn{
% \begin{itemize}
%     \item $E_\mathcal{H}$, an enumeration of desugarings
%     \item $\mathcal{I}$, a test set (list of source programs)
%     \item $P_\Sigma$, an array of outputs for programs in the test set by the source interpreter \\ (i.e. $P_\Sigma[i]$ is the output for input $i$ by the source interpreter)
%     \item $\Phi_\Omega$, an opaque interpreter for the core language
% \end{itemize}
% }
% \KwOut{A correct desugaring $\delta$, or timeout}
% \SetKwFunction{search}{SearchDesugaring}
% \SetKwFunction{test}{TestDesugaring}
% \SetKwProg{function}{Function}{}{}
% \function{\search{}}{
%   \ForAll {$\delta \in E_\mathcal{H}$}{ 
%      \If {TestDesguaring($\delta$)}{
%       \Return $\delta$
%      }
%   } or \emph{Timeout} }{}
% \function{\test{$\delta$}}{
%      $S_\Sigma \longleftarrow \emptyset , S_\Omega \longleftarrow \emptyset$ \\
%       \ForAll {$i \in \mathcal{I} $}{
%             \If {$\delta(P_\Sigma[i]) \neq \Phi_\Omega(\delta(i))$}{
%                 \Return False
%             }
%             add $P_\Sigma[i]$ to $S_\Sigma$ (set addition)\\
%             add $\Phi_\Omega(\delta(i))$ to $S_\Omega$ (set addition)
%      }
%      \If {$|S_\Sigma| \neq |S_\Omega| $} {
%           \Return False
%      }
%      \Return True
%   }{}
%   \caption{Desugaring search}\label{alg:desugaring}
% \end{algorithm}

The algorithm sequentially tests each desugaring of the enumeration $E_\mathcal{H}$ until a correct one is found or we reach a timeout. The test condition directly corresponds to our definition of correctness with respect to the test set $\mathcal{I}$: we test soundness for every test program, and test adequacy by testing whether the number of distinct outputs (values or errors) we get by either the source or the target interpreter is equal. If they are, this, together with the already tested soundness, ensures the injectivity of the translation on the values obtained from the test set.

The main complexity of this simple algorithm lies in the definition (and efficient implementation) of the enumeration of desugarings $E_\mathcal{H}$. Indeed, the main point of the algorithm is to reduce the problem to an enumerative search: to the definition of a hypothesis space and an enumeration of that hypothesis space. We might choose different hypothesis spaces for different desugaring extension sub-problems, based on domain knowledge.

In the rest of the section we define various hypothesis spaces of increasing expressiveness.  The hypothesis spaces (sets of interpretations) are defined inductively, parametrised by a given source signature $\sigma_1 \times \dots \times \sigma_n \to \sigma$. The full translation is generated by a tuple of such interpretations for each term constructor in $\Sigma \setminus \Sigma'$. We will not investigate different enumeration strategies: our enumerations are the same that we would get by breadth-first search (but we use a more efficient implementation, to be discussed in Section ~\ref{sec:implementation}).

\subsection{Relabelling} The simplest hypothesis space we will consider, $\mathcal{H}_{\text{relabel}}$, is formed by the term constructors of the target language. With $\mathcal{H}_{\text{relabel}}$ we can express desugarings which map each constructor of the source language to a term constructor of the target language of the same signature: a relabelling. Note that any arguments of the source term constructor are translated recursively and the order of arguments cannot be changed.  We will call the desugaring extension problem specialised to $\mathcal{H}_{\text{relabel}}$ the \emph{relabelling problem}.

In our intended translation many rules can be expressed as relabellings: in fact all of them except but $\src{\id{STrue}}$, $\src{\id{SFalse}}$, $\src{\id{SPrim}}$, $\src{\id{SBetween}}$, and $\src{\id{SFor}}$ can:
\[\begin{array}{rcl}
    \sem{ \id{SNum}(n) } &= &\cor{\id{CNum}(\sem{ n })} \\
    \sem{ \id{SVar}(n) } &= &\cor{\id{CVar}(\sem{ n })} \\
    \vdots & \vdots & \vdots \\
%    \sem{ \id{SStr}(n) } &=& \cor{\id{CStr}(\sem{ n })} \\
%    \sem{ \id{SIf}(t_1,t_2,t_3) } &=& \cor{\id{CIf}( \sem{  t_1  } , \sem{  t_2  } , %\sem{  t_3  } ) } \\
%    \sem{ \id{SLam}(is,t) } &=& \cor{\id{CLam}(  is  , \sem{  t  } ) } \\
%    \sem{ \id{SApp}(t_1,t_2) } &= &\cor{\id{CApp}( \sem{  t_1  } , \sem{  t_2  } ) } \\
%    \sem{ \id{SLet}( i , t_1 , t_2 ) } &=& \cor{\id{CLet}(  i  , \sem{  t_1  } , %\sem{  t_2  } ) } \\
%    \sem{ \id{SLetRec}(i,t_1,t_2) } &= &\cor{\id{CLetRec}( i  , \sem{  t_1  } , %\sem{  t_2  } ) } \\
%    \sem{ \id{SAssign}(i,t) } &=& \cor{\id{CAssign}(  i  , \sem{  t  } ) } \\
%    \sem{ \id{SList}(\bar{t}) } &=& \cor{\id{CList}( \sem{ \bar{t} } ) } \\
   \sem{ \id{SListCase}(t_1,t_2,t_3) } &= &\cor{\id{CListCase}( \sem{ t_1 } , \sem{  t_2  } , \sem{  t_3  } ) }
\end{array}\]

Of course there are also translation rules in $\mathcal{H}_{\text{relabel}}$ that are not correct (and therefore not our intended translation):
\begin{align*}
    \sem{ \id{SLetRec}(i,t_1,t_2) } &= \cor{\id{CLet}(  i  , \sem{  t_1  } , \sem{  t_2  } ) }
\end{align*}

This example shows that even if we assume that the intended translation is a relabelling we still may need appropriate test cases that rule out incorrect translations, and they are not always completely trivial to find. To distinguish between the recursive and non-recursive \lstinline{let} construct we need an example which actually behaves recursively when interpreted as \lstinline{letrec}.

\subsection{Substitution} 

The next hypothesis space, $\mathcal{H}_{\text{subst}}$, is defined by terms with variables and substitution.

\begin{definition}[Terms with variables] Let $T_\Omega$ be an abstract language over signature $\Omega(S,\mathcal{F},\sigfun)$, where $S = \{ \sigma_1, \dots, \sigma_n \}$. Let a function $\rho : S \to \mathbb{N}$ assign a natural number to every sort in the signature. Let $X^\rho$ be a finite set of variables, containing $\rho(\sigma)$ variables for every sort $\sigma$:
$$X^\rho = \{ x^{\sigma_1}_1,\dots,x^{\sigma_1}_{\rho(\sigma_1)},\cdots,x^{\sigma_n}_1,\dots,x^{\sigma_n}_{\rho(\sigma_n)} \} $$
%, and let $X$ be 
and disjoint from $\mathcal{F}$. Extend the signature function to the new variables as
$$ \forall \sigma \in S, \forall i \in [ 1 \dots \rho(\sigma) ], \sigfun(x^\sigma_i) = \sigma $$

We denote the the set of terms with variables from $X^\rho$ as $T_\Omega(X^\rho)$. The terms of $T_\Omega(X^\rho)$ are called \emph{terms with variables} over the language $T_\Omega$. A term $t_{\mathcal{X}^\rho} \in T_\Omega(\mathcal{X}^\rho)$ defines a function by substituting the arguments into the variables. We will refer to translations (and interpretations) that can be expressed by terms with variables as \emph{substitutions}.
\end{definition}

Let $f \in \mathcal{F}_\Sigma$, and let $\rho_f(\sigma)$ be the number of times $\sigma$ occurs in the domain of the signature of $f$. The hypothesis space is defined as $\mathcal{H}^f = T_\Omega(\mathcal{X}^{\rho_f})$.

Every relabelling is a substitution. The resulting task is close to the formulation of \KLE, who suggested modelling desugarings as tree transducers~\cite{tata2007}. The translations defined by translation rules in $\mathcal{H}_{\text{subst}}$ are also definable by a deterministic top-down tree transducer: $\mathcal{H}_{\text{subst}}$ guarantees that the output of a translation is well-typed according to the core language's signature (corresponds to the grammar), and that the translation preserves the sorts. 

A further difference is that in our hypothesis space a term constructor determines the translation rule. The two translation rules for $\src{SPrim}$ can not be expressed in this hypothesis space. Expressing the two rules for $\src{SPrim}$ is possible with a bottom-up tree transducer or a non-deterministic top-down tree transducer (both of which are more expressive than deterministic top-down tree transducers), but requires additional states that do not correspond to the sorts defined by the grammar. 

While this hypothesis space is much more expressive than relabellings, only two additional rules of the example intended translation can be covered:
\begin{align*}
   \sem{ \id{STrue}  } &= \cor{\id{CBool}(\mathit{true})}  & 
    \sem{ \id{SFalse} } &= \cor{\id{CBool}(\mathit{false})} 
\end{align*}

%\id{SBetween} can almost be covered, but the fresh variable generation needed cannot be defined as a substitution.
The main reason is that the Pidgin source and core languages were specifically chosen by \KLE to show potential problems with modelling the translation with a tree transducer. In the following section, we introduce one of our main contributions, a simple model which can express all of Pidgin's translation rules, as well as the multiple rules for $\src{\id{SPrim}}$.

\subsection{Meta-program templates}

In general, the translations could be written in an appropriate meta-language, which could easily express all intended translation rules. But synthesising translation rules in a general purpose language is a very challenging task: we are looking for the most restricted hypothesis space that can still express the translations. As a middle ground, we consider extending the former hypothesis space $\mathcal{H}_{\mathrm{subst}}$ with a fixed finite set of templates, written in a general-purpose meta-language that allows expressing the remaining cases. 

\begin{figure}[tb]
    \centering\small
    \[\small\begin{array}{rcl}
    \sigma_1,\dots,\sigma_n,\sigma &\in &S \\
    f &\in& \mathcal{F}\\
    t &::=& x \mid c \mid f(t_1,\ldots,t_n)
    \end{array}\]
    \begin{displaymath}
    \prftree[r]{\textsc{Axiom}}{\Gamma ; x : \sigma \vdash x : \sigma} \qquad \prftree[r]{\textsc{C-rule}}{c : \sigma}{\Gamma \vdash c : \sigma}
   \qquad
    \prftree[r]{\textsc{F-rule}}{\Gamma \vdash t_1 : \sigma_1}{\cdots}{\Gamma \vdash t_n : \sigma_n}{f : \sigma_1 \times \dots \times \sigma_n \to \sigma}{\Gamma \vdash f(t_1,\dots,t_n) : \sigma}
    \end{displaymath}
    \caption{Terms with variables as proofs}
    \label{fig:base_sequent_rules}
    \[\small\begin{array}{rcl}
    t &::= & \dots \mid \mathbf{let}\, i = \mathtt{gensym}() \, \mathbf{in} \, \cor{\id{CLet}}(\lit{i}, t_1 , t_2) \\
    M, N &::= & t \mid \mathbf{let}\, (x_1,x_2) = \mathtt{unzip}(x) \,\mathbf{in}\, M \mid \mathbf{syntax\_error} \mid \mathbf{case}\, y \,\mathbf{of}\, [] \to M ; (x:y) \to N : \sigma'  
    \end{array}\]
    
\~\
    
\begin{displaymath}
\prftree[r]{\textsc{Unzip}}{\Gamma ; x_1 : [\sigma_1] , x_2 : [\sigma_2] \vdash M : \sigma}{\Gamma ; x : [\sigma_1 \times \sigma_2] \vdash \mathbf{let}\, (x_1,x_2) = \mathtt{unzip}(x) \,\mathbf{in}\, M : \sigma}
\qquad
\prftree[r]{\textsc{Throw}}{\Gamma \vdash \mathbf{syntax\_error} : \sigma}
\end{displaymath}
\\
\begin{displaymath}
\prftree[r]{\textsc{Case}}{\Gamma \vdash M : \sigma'}{\Gamma ; x : \sigma , y : [\sigma] \vdash N : \sigma'}{\Gamma ; y : [\sigma] \vdash \mathbf{case}\, y \,\mathbf{of}\, [] \to M ; (x:y) \to N : \sigma'}
\end{displaymath}
\\
\begin{displaymath}
\prftree[r]{\textsc{Fresh}}{\Gamma \vdash t_1 :\mcor{\sort{CTerm}}}{\Gamma ; x :\mcor{\sort{CTerm}} \vdash t_2 :\mcor{\sort{CTerm}}}{\Gamma  \vdash \mathbf{let}\, i =  \mathtt{gensym}() \, \mathbf{in} \, \cor{\id{CLet}}(\lit{i},\lit{t_1},\lit{t_2}[\mcor{\id{CVar}}(i)/x]):\mcor{\sort{CTerm}}}
\end{displaymath}

    \caption{Rules for meta-program templates}
    \label{fig:extended_sequent_rules}
\end{figure}

In our current approach, it is up to the user to determine the set of templates used at an individual learning step. Our contribution is a general framework (implemented as a library) that can incorporate various templates and provide an enumeration of all well-typed translations formed by them. %a set of templates.
We demonstrate the technique with a set of templates we wrote for the Pidgin languages.
  
To formally describe the new system $\mathcal{H}_{\text{meta}}$, we start with the re-formulation of $\mathcal{H}_{\text{subst}}$ as deductive rules, in a sequent calculus style, shown in Figure \ref{fig:base_sequent_rules}. Our judgements describe meta-language programs. They are of the form $\Gamma \vdash t : \sigma$, where $\Gamma$ is the context (variable declarations, the argument names and their sorts), $t$ is a term of the meta-language describing a translation rule, and $\sigma$ is a sort of the output. The context is a set of pairs of variables with their corresponding sort in the form of  $x : \sigma$, where $x$ is a meta-variable name (not an identifier in the source or core languages), and $\sigma$ is a sort. $\Gamma$ will range over such contexts, and we extend a context $\Gamma$ with additional variable declarations $x : \sigma_1$ and $y : \sigma_2$ as $\Gamma ; x : \sigma_1 , y : \sigma_2$. We identify $\alpha$-equivalent meta-terms. In the implementation the meta-variable names in the context $\Gamma$ are determined by the signature (the set of sorts), but in the presentation we avoid this complication. Note that this only concerns the meta-variables, and not any identifiers (if present) in the core language: we did not assume to know the $\alpha$-equivalence relation of the core language (if any).

Enumerating all translation rules corresponding to a source signature $\sigma_1 \times \dots \times \sigma_n \to \sigma$ is thus implemented as enumerating all proofs of the judgement $x_1 : \sigma_1 , \dots , x_n : \sigma_n \vdash t : \sigma$. This way we reduce the desugaring extension problem to enumerative synthesis, that allows us to compare various hypothesis spaces and enumeration strategies.

We could have chosen different styles of sequent calculus to describe the same system. Our style has the big advantage that proofs and terms (with variables) directly correspond to each other: there are no permutations of the proof rules that yield effectively the same terms, thus enumerating the terms is much easier. 

Now that we have expressed our translations as proof terms, we can extend them with arbitrary templates from a functional meta-language. Figure \ref{fig:extended_sequent_rules} shows some possible meta-program templates added as derivation rules. We use two kind of templates.

The first group are meta-programs that act on the source terms: we call them left-rules since they operate on the left side of the judgement. We may note that some sorts are structurally equivalent to lists or tuples of other sorts, and use them as such in the meta-programming language. For example, we introduced $\texttt{List}_{\msrc{\sort{SForBind}}}$ for a list of terms belonging to the sort $\msrc{\sort{SForBind}}$, while the sort $\msrc{\sort{SForBind}}$ itself is structurally equivalent to a tuple of an identifier and a source term. This allows us to use functions like $\mathtt{unzip}$. In the presentation we used the notation $[\sigma]$ to mean a sort that is structurally equivalent to a list, and $\sigma_1 \times \sigma_2$ to mean a sort that is structurally equivalent to a tuple. The first rule, \mr{Unzip}, turns a list of tuples into two lists. With this primitive operation we can express the rule for $\src{\id{SFor}}$. The \mr{Case} rule allows us to destruct a list, and allows us to express the two rules for $\src{\id{SPrim}}$ (distinguished by the number of children) as one. But the \mr{Case} rule alone is not sufficient, as we need to do something when the list of children is empty, which is allowed by our source grammar, but results in an error in the source language interpreter. To represent compile-time errors not caught by the grammar we introduce a meta-program for syntax errors. Note that this assumes that translation rules are performed in the error monad: compile time errors are observable side-effects, and we need a way to express them.

The second group are meta-programs that encode typical core program templates. This technique is currently demonstrated by a single rule: the \mr{Fresh} rule demonstrates fresh name generation, which allows us to express the translation rule for $\src{\id{SBetween}}$. The meta-function $\mathtt{gensym}()$ generates unique identifiers for the core language, which are bound to a meta-language variable $i$. We use this fresh variable as the local variable introduced by \cor{\id{CLet}}, and in the scope of the variable we replace the meta-argument with dereferencing it. This rule relies on additional domain knowledge: the binding and scoping information of the core language (but can be auto-generated if this information is available in addition to the grammar). Note that in the scope we only allow dereferencing the identifier as a variable ($\mcor{\id{CVar}(\lit{i})}$), not to use it in arbitrary positions identifiers may occur, for example as a local variable introduced by a $\cor{\id{CLam}}$.

Using $\mathcal{H}_{\mathsf{meta}}$, we can express the remaining desugaring rules.
\[\small\begin{array}{rcl}
\sem{ \id{SBetween}(t_1,t_2,t_3) } &=& \mathbf{let}\, i_1 = \mathtt{gensym}() \, \mathbf{in}\, \cor{\id{CLet}(i_1, \sem{ t_1 }}, \\
&&\mathbf{let}\, i_2 = \mathtt{gensym}() \, \mathbf{in}\, \cor{\id{CLet}(i_2, \sem{t_2}},\\
&& \mathbf{let}\, i_3 = \mathtt{gensym}() \, \mathbf{in}\, \cor{\id{CLet}(i_3, \sem{t_3} , } \\ && \cor{ \id{CPrim2}(\land,\id{CPrim2}(<,  \id{CVar}(i_1) , \id{CVar}(i_2) ), \id{CPrim2}(<, \id{CVar}(i_2)  ,  \id{CVar}(i_3)  )))))}\\\
    \sem{ \id{SPrim}(\lit{o},ts) } &=& \mathbf{case}\, \sem{ts} \,\mathbf{of}\, [] \to \mathbf{syntax\_error} ; \\
    &&\qquad \qquad (t_1:ts_1) \to \mathbf{case}\, ts_1 \,\mathbf{of}\, [] \to \cor{\id{CPrim1}( \lit{  o  } ,   t_1 ) } ; \\
    &&\qquad \qquad \qquad \qquad \qquad \qquad (t_2:ts_2) \to \mathbf{case}\, ts_2 \,\mathbf{of}\, [] \to \cor{\id{CPrim2}( \lit{  o  } ,   t_1 , \ t_2  ) } ; \\
    &&\qquad \qquad \qquad \qquad \qquad \qquad \qquad \qquad \qquad \qquad (\_:\_) \to \mathbf{syntax\_error} \\
    \sem{ \id{SFor}(t_1,bs,t_2) } &=& \mathbf{let}\, (ids,ts) = \mathtt{unzip}(\sem{bs}) \,\mathbf{in}\\
    &&\cor{ \id{CApp}( \sem{ t_1 } , [ \id{CLam}(\lit{ids} , \sem{t_2}) , \id{CList}(\lit{ts}) ] ) }
\end{array}\]

Note that we could have included a more general rule instead of Fresh:

\begin{displaymath}
\prftree[r]{\textsc{Fresh$'$}}{\Gamma ; y : \id{Id} \vdash M :\mcor{\sort{CTerm}}}{\Gamma \vdash \mathbf{let}\, y = \mathtt{gensym}() \, \mathbf{in} \, M :\mcor{\sort{CTerm}}} 
\end{displaymath}

This version only relies on $\mathtt{gensym}$ being a meta-function generating identifiers. It is possible to express the intended translation with the more general \mr{Fresh$'$} rule. This version, however, blows up the search space as fresh names could be generated anywhere. In our scope-aware version we relied on the domain knowledge that fresh names only need to be introduced in language constructs that bind new names, and only allowed to be used as variable references in the scope of the binding. This can be extended to other name-binding constructs, and \mr{Fresh} can be seen as an example of a general pattern: we will show two other examples of it in Section \ref{sec:evaluation}.

But introducing the meta-rules breaks the nice property of the correspondence between proofs and translations: there are multiple proofs corresponding to essentially the same translation, because the left rules such as \mr{Unzip} or \mr{Case} can be permuted with the right rules and the core constructors. This prevents us to efficiently enumerate such programs with our simple enumeration algorithm, and results in the explosion of the search space. 

We remedy the problem by restricting the order of the templates by layering. Note that the left rules can always be pulled to the top of the term: in the meta language the left rules are functions while the rest are data. (To be precise the \mr{Fresh} rule contains both, but it does not affect the uniqueness of the translation.) We divide the rules to right rules (the \mr{Axiom}, the \mr{C-rule}, the \mr{F-rule} and the \mr{Fresh} rule) and left rules (\mr{Unzip}, \mr{Case}, and \mr{Throw} rules), and require that left rules are never used in the sub-terms (premises) of a right rule. This layering does not completely solve the problem: there still could be permutations amongst the left rules that lead to identical translations, but it is hard to exclude them without analysing the rules.

As a summary, we may speculate how much user assistance is needed to set up the necessary meta-rules. We can identify three general patterns: to deconstruct source terms (\mr{Case}), to allow compile time errors (\mr{Throw}) and to generate fresh names in binding positions (\mr{Fresh}). We can imagine these meta-rules to be auto-generated. But our last meta-rule, \mr{Unzip}, does not completely fall into these patterns, and shows that in some cases more user assistance is needed.

\subsection{Restricting and combining hypothesis spaces}

The last hypothesis space, $\mathcal{H}_{\text{meta}}$, contains every intended translation rule. But this hypothesis space is too large for our simple enumerative algorithm, and in some test cases we failed to learn the intended desugaring rule within our timeout of 1 hour on the test machine. We introduce two techniques, layering and restriction, to define subsets of $\mathcal{H}_{\text{meta}}$.

We have already used layering to prune some equivalent translations from the enumeration. We allow the user to specify additional layers. One natural additional layer is to only allow the \mr{Fresh} rule (and similar fresh identifier introducing rules) at the top (below left rules, but above core term constructors). The reasoning is similar: it is likely that these rules would be at the top if we could convert core language terms to $\beta$-normal form. However, unlike the meta-language, the core language is a parameter of the problem, and in general we do not know its equivalence relation. Therefore we handle the layer for the left rules as fixed, while leaving introducing a layer for the \mr{Fresh} rule in the hands of the user, who can leverage domain knowledge of the core language.

Layering alone is still not enough, and another straightforward technique to define a subset of $\mathcal{H}_{\text{meta}}$ is to only allow a subset of rules. We allowed the user to restrict meta-rules to a subset, relying on domain knowledge. This is not surprising: a typical core language, even our simplified Pidgin core, is much larger than the DSLs targeted by successful program synthesis methods. 

In the desugaring extension problem we may need to search for the translation rules of multiple term constructors at the same time. We currently always use fair interleaving: without domain specific knowledge about the languages we can not assume a bias which will generalise well.

\section{Implementation} \label{sec:implementation}

We used Haskell to carry out our investigations. Writing test cases requires implementing source and core languages, translations between them, enumerations of such translations, and implementing the search algorithm.  

The implementation of the small source and core languages involved in the tests as embedded languages is standard 
%Uncomment to build TR
(simplified versions of these implementations are shown in Appendix \ref{appendix:reference_impl}). 
%(simplified versions of these implementations are shown in the companion technical report~\cite{bartha21arxiv}). 
To deal with non-terminating programs we implemented bounded evaluation (limiting the number of steps allowed in the interpreter).

The implementation of the simple linear search over the enumerations based on the testing framework shown in Figure \ref{fig:testing} is also straightforward. The crucial part of the implementation, which we discuss in some detail, is the implementation of the enumeration of the translations, in particular, how to enumerate efficiently the proofs of $\mathcal{H}_{\text{meta}}$.

We have implemented a library that can efficiently enumerate the proof terms generated by arbitrary rules corresponding to our sequent calculus-like formalism. The user of the library can write  arbitrary Haskell functions that either operate on the source language terms or serve as templates for core language terms, corresponding to our two types of meta-rules%, and incorporate them in the search
. The sequent calculus rules ensure that the library only enumerates well-typed terms.

We tested depth-first search and iterative deepening, but they proved to be too memory intensive and slow. Our current solution builds on the FEAT Haskell library~\cite{Duregard:2012}, that provides efficient functional enumerations of algebraic data types (ADTs) out of the box. 

Proofs of judgements of the form $x_1:\sigma_1,\ldots,x_m:\sigma_n \vdash T : \sigma$ can not be represented naturally by an ADT. We want to restrict meta-variables to those that are available in the environment $x_1:\sigma_1,\ldots,x_m:\sigma_n$, and  we would need a separate ADT for each environment in order to enforce this. But meta-rules can introduce new environments. We implemented environment-dependent enumerations relying on the library primitives, and we used memoisation and cached the generated enumerations for the given environment. The enumerations are recursive in a non-trivial way and our implementation quickly exhausted available memory without this optimisation. 

The enumerative search is embarrassingly parallel, and we expect a fully parallel implementation to help a lot. So far we only did a preliminary investigation of parallelism relying on Haskell's deterministic parallelism capabilities \cite{Marlow:2009}, and while it clearly speeds up the search, most likely better performance could be achieved with a more fine tuned parallel implementation. Our experiments report timings with deterministic parallelism employed. In recent versions of GHC the garbage collector can exploit multiple cores: we also enabled this optimisation for the compiler. The code to reproduce our case studies is available at \url{http://10.5281/zenodo.5475211}.

\section{Evaluation} \label{sec:evaluation}

We evaluate our approach and the enumeration synthesis library on three case studies. The first is the Pidgin languages that we used to demonstrate our approach throughout the paper. As a preliminary exploration of how well the method generalises, as well as to demonstrate some limitations of our approach, we extend the Pidgin languages with two sets of language constructs: list comprehensions and exception handling. We extend both the core language and the source language, and investigate whether we can extend our sequential learning algorithm to the new cases seamlessly.  Our experiments were run on a Intel E3-1245v5 @ 3.50GHz with 8 cores and 32 GB RAM, running Debian Linux 10 and GHC 8.8.4.

\subsection{Hypothesis}

Our hypothesis is that for the challenge problem we can set up a sequential learning task where each individual desugaring extension problem can be solved with our enumerative synthesis algorithm and combined to solve the overall problem. Moreover we also hypothesise that when we extend the core and source languages with new constructs we can extend the sequential learning task with new steps as well, and solve these new desugaring extension problems with the same algorithm. 

As the hypothesis space is a parameter of the algorithm (and the desugaring extension problems) and ultimately in the hand of the user, it is hard to measure how well the method generalises: on one hand, we can set up a hypothesis space that only contains the intended translation as a meta-rule, and on the other hand we could fill the hypothesis space with so many meta-rules that enumerative search is hopeless. Our method was to start with a (to us) natural hypothesis space, and in the cases where we were not able to get results within a 1 hour timeout, we repeated the test with additional user assistance. While we do report various metrics from our tests, we think the amount of user assistance needed for each step to successfully complete is the most useful metric for an envisioned semi-automated assistant.

\subsection{Pidgin languages}

\textit{Setup.}
The Pidgin source and core languages was shown in Figure \ref{fig:src_and_core}. We assumed that part of the translation is known in advance. The sublanguage $\Sigma_0$ contains numbers, strings, identifiers and the constructors for lists, and we assumed that their translation is known (which means that numbers, strings and identifiers and list of identifiers are preserved, and lists of source terms are translated to lists of core terms by individually translating the elements).

\textit{Hypothesis space.}
The starting hypothesis space $\mathcal{H}^1$ contained all core term constructors, including list constructors for term and identifier lists. It did not contain number, string or identifier constants. In general, we do not want to allow such constants in the hypothesis space: they rarely occur in desugaring rules and they blow up the search space. However, we need the Boolean constants \texttt{true} and \texttt{false}, which is clear, since the Boolean sort is not part of the source language. We also included all operators, which requires some domain knowledge about their role: while operators are constants, they correspond to various functions. The hypothesis space also contained all meta-rules listed in Figure \ref{fig:extended_sequent_rules}: \mr{Case}, \mr{Throw}, \mr{Unzip} and \mr{Fresh}.

We also use a second, specialised hypothesis space $\mathcal{H}^2$. This hypothesis space only allowed the core term constructors $\cor{\id{CVar}}$ and $\cor{\id{CPrim2}}$, the meta-rule \mr{Fresh}, and all constants from $\mathcal{H}^1$, reducing the size of the hypothesis space significantly.

\textit{Results.}
We run the first test with $\mathcal{H}^1$, and we found all intended translation rules within the time limit but one: this test did not find the intended desugaring for  $\src{\id{SBetween}}$.
% $$ \begin{array}{rcl}
% \sem{ \id{SBetween}(t_1,t_2,t_3) } &=& 
%     \cor{\id{CLet}(\%i_1, \sem{ t_1 }, \id{CLet}(\%i_2, \sem{t_2}, \id{CLet}(\%i_3, \sem{t_3} , } \\ && \cor{ \id{CPrim2}(\land,\id{CPrim2}(<,  \%i_1 , \%i_2 ), \id{CPrim2}(<, \%i_2  ,  \%i_3  )))))} \end{array}$$
    
% Instead, it found an expression that may evaluate $t_2$ twice:
% $$ \sem{ \id{SBetween}(t_1,t_2,t_3) } = 
%     \cor{\id{CPrim2}(\land, \id{CPrim2}(<, \sem{  t_1  } , \sem{t_2} ), \id{CPrim2}(<, \sem{  t_2  } ,  \sem{  t_3  }  )}$$
% Note that we only found this simplified rule because we did not include test cases for $\src{\id{SBetween}}$ that depend on the times $t_2$ is evaluated: if we include the additional test cases that ensure that the proper translation is found, this learning step timeouts (1 hour timeout on the test machine).

We run a second test where we replaced the desugaring extension problem for $\src{\id{SBetween}}$ (step $\Sigma_7$) based on $\mathcal{H}^1$ with a new one based on $\mathcal{H}^2$.
% and additional source programs whose result depends on the order and number of times each argument is evaluated.
The new sequential learning task gives us a full solution of the challenge problem. It is summarised in Table~\ref{tab:pidgin_case_study}.

\begin{table}[h]
\small
\begin{tabular}{l | c | c | c | c | c | l}
Task & New constructors & Hyp. space & AST size & Index & \#test set & Time \\
\hline \hline
$\Sigma_1$ & $\src{SNum}$ & $\mathcal{H}^1$ & 2 & 1 & 2 & 0s \\
$\Sigma_2$ & $\src{SStr}$ & $\mathcal{H}^1$ & 2 & 1 & 2 & 0s \\
$\Sigma_3$ & $\src{SPrim}$ & $\mathcal{H}^1$ & 12 & 16567100 & 5 & 3min9s \\
$\Sigma_4$ & $\src{SVar}, \src{SLet}$ & $\mathcal{H}^1$ & 6 & 1298 & 2 & 0s \\
$\Sigma_5$ & $\src{STrue}, \src{SFalse}$ & $\mathcal{H}^1$ & 4 & 10 & 4 & 0s \\
$\Sigma_6$ & $\src{SAssign}$ & $\mathcal{H}^1$ & 3 & 18 & 1 & 0.0s  \\
$\Sigma_7$ & $\src{SBetween}$ & $\mathcal{H}^2$ & 14 & 157160392 & 9 & 28min2s \\
$\Sigma_8$ & $\src{SIf}$ & $\mathcal{H}^1$ & 4 & 171 & 2 & 0s \\
$\Sigma_9$ & $\src{SLam}, \src{SApp}$ & $\mathcal{H}^1$ & 6 & 1813 & 2 & 0s \\
$\Sigma_{10}$ & $\src{SLetRec}$ & $\mathcal{H}^1$ & 4 & 132 & 1 & 0s \\
$\Sigma_{11}$ & $\src{SList}$ & $\mathcal{H}^1$ & 2 & 3 & 2 & 0s \\
$\Sigma_{12}$ & $\src{SListCase}$ & $\mathcal{H}^1$ & 4 & 207 & 2 & 0s \\
$\Sigma_{13}$ & $\src{SFor}$ & $\mathcal{H}^1$ & 11 & 57334664 & 3 & 9min29s  \\
$\Sigma$ & total             &&&& 35 & 40min40s
\end{tabular}
\caption{Case study -- Pidgin languages}
\label{tab:pidgin_case_study}
\end{table}

The rows of the table correspond to desugaring extension problems. In each problem $\Sigma_n$ the translation of the sublanguages $\Sigma_0, \dots, \Sigma_{n-1}$ are assumed to be known. We work our way downwards: the test sets can not use term constructors from below. To get some rough measurement of the complexity of each task we noted the size of the intended desugaring (AST size), and also the \emph{index} of the translation found in the enumeration for each desugaring extension problem, that is, the number of attempts tried before a solution was found. We also show the number of handwritten tests we used to find the solution. The solution found was exactly the intended translation in Figure \ref{fig:src_and_core} in all but one case: $\src{\id{SBetween}}$. 

In the case of $\src{\id{SBetween}}$ the algorithm finds an expression equivalent to our intended translation but smaller:
\begin{displaymath}
\begin{array}{rcl}
\sem{ \id{SBetween}(t_1,t_2,t_3) } &=& 
    \cor{\id{CLet}(\%i_1, \sem{ t_1 }, \id{CLet}(\%i_2, \sem{t_2}, }\\ && \cor{ \id{CPrim2}(\land,\id{CPrim2}(<,  \%i_2, \sem{t_3} ), \id{CPrim2}(<, \%i_1  ,  \%i_2  )))))} \\
    
\end{array}
\end{displaymath}

\textit{Decomposition.}
The desugaring extension problems' test sets can not contain source term constructors from later extensions, but the test set still needs to exclude all non-intended translations. This means that we can not learn $\src{SLam}$ and $\src{SApp}$ separately, as they only show their behaviour together: we can see that they are grouped together in the task $\Sigma_4$.

\KLE did not specify the exact semantics of the source and core languages --- this give us a little bit of freedom implementing them. Our implementation does not allow executing open programs containing non-declared variables, that is, all variables must be declared in a $\src{SLam}$, $\src{SLet}$ or $\src{SLetRec}$ expression before use, otherwise an exception is raised. This means that we can not use $\src{SVar}$ on its own without one of these: we grouped it together with $\src{SLet}$ in task $\Sigma_9$.

The last case where we needed to group multiple term constructors together was the $\src{STrue}$, $\src{SFalse}$ group. The symmetry of the Boolean constants means that we need to learn them together and rely on the already fixed translations of the \texttt{and} and \texttt{or} operations to rule out translations that map to the opposite Boolean value in the core language.

Since all other groups only contain one term constructor, this decomposition shows that quite often we do not need to learn more than one rule at a time. This supports our hypothesis that the language can be partitioned into small groups of term constructors.

\textit{Test cases.}
The majority of test programs are very small and simple: in most cases we do not need large test programs, and it is plausible that they could be generated by automatic testing. There are three exceptions. To learn the full semantics of $\src{\id{SBetween}}$ we needed test programs where the evaluation order of arguments matters. To distinguish $\src{SLetRec}$ from $\src{SLet}$ we needed recursive examples. The source term constructor $\src{SFor}$ assumes a combinator function as its first argument, which leads to an example that dwarfs the rest. This last case is somewhat artificial, though.

In many cases our algorithm found a desugaring that is equivalent to the intended semantics on pure terms, but differs in the evaluation order of the arguments. Pidgin source has side effects: the $\src{\id{SAssign}}$ operator. We utilised it to add test cases that depend on the evaluation order.

%One important remark about the test programs is the need for examples that depend on the evaluation order. Pidgin source has side effects: the $\src{\id{SAssign}}$ operator. In many cases our algorithm found a desugaring that is only equivalent to the intended semantics on pure terms, but differs in the order of the evaluation of the arguments. We needed to write new examples that disambiguate the desugaring rule.

Sometimes the order of the examples is important. The majority of the time is spent evaluating various core programs that we get by the tested translations, especially when -- like in the case of $\src{SFor}$ -- we need recursive examples. Recursive examples easily lead to non-terminating core programs, thus the time spent checking prospective translations on them greatly depends on the maximum number of steps allowed in the interpreter. If this bound is too low, we may get an incorrect desugaring, if too high, the search may take a long time.  We can learn the intended semantics of $\src{SFor}$ with only one test program, but since it needs to be recursive and large, the search time suffers, and depends on the value of the parameter making the learning process impractical. As currently we can only tune this parameter manually (automatic tuning is left to future research), it is important for performance to use simple, non-recursive source programs as the first test cases in a test suite. With our test setup (in which we added such a test program before the full example) we found that a reasonably high bound parameter does not have a large effect on the time of the search, so no manual tuning was needed, while the search time was reduced by around 90\%.

\subsection{Basic list comprehensions}

\textit{Setup.}
For our first extension of the Pidgin languages we consider basic list comprehensions a la \citet{wadler_1992} with the following syntax: 
\[\small
  \begin{array}{rclcrcl}
  t &::=& \cdots \mid [t \mid q]&&
q &::=& \epsilon \mid x \leftarrow t, q \mid t, q \mid \mathrm{let}~x = t, q
\end{array}\]
This is a simplification over standard Haskell list comprehensions, which support patterns in bindings and more general definitions in \verb|let|.
The following table shows the extensions of the syntax of the Pidgin languages required for this case study:

\begin{minipage}{\linewidth}
\centering
\[\small
  \begin{array}{rcl}
    t  \in \msrc{\sort{STerm}} &::= & \dots \bnfor
    \msrc{ \id{SListComp}(t,q)} \\
    q \in \msrc{\sort{SQual}} &::=&
    \msrc{ \id{QEmpty} } \bnfor
    \msrc{ \id{QBind}(\lit{i}, t, q) } \bnfor                               
    \msrc{ \id{QGuard}(t,q) } \bnfor
    \msrc{ \id{QLet}(\lit{i},t,q) }\\
          e, e_1 , e_2  \in \mcor{\sort{CTerm}} &::= & \dots \bnfor
    \mcor{ \id{MVar}(\lit{i}) } \bnfor
    \mcor{ \id{MLam}(\lit{i},\lit{e}) } \bnfor
    \mcor{ \id{MApp}(\lit{e_1},\lit{e_2}) } \bnfor
    \mcor{ \id{Return}(e) }\bnfor
    \mcor{ \id{Bind}(e_1,e_2) }%\bnfor
%    \mcor{ \id{Seq}(e_1,e_2) }\bnfor
%    \mcor{ \id{Guard}(e) }
\end{array}\]
\end{minipage}

The $\mcor{\id{Return}}$ and $\mcor{\id{Bind}}$ primitives correspond to the following standard Haskell functions:
\begin{verbatim}
return :: a -> [a]   (>>=)  :: [a] -> (a -> [b]) -> [b]
\end{verbatim}
The desugaring of comprehensions is described as a two-argument function $D[t | q]$ in the GHC documentation\footnote{\url{https://ghc.gitlab.haskell.org/ghc/doc/users_guide/exts/monad_comprehensions.html})}.  This does not quite fit our framework, but can be expressed by desugaring qualifiers to functions $\sem{q}$ mapping core terms to core terms, and desugaring a comprehension by applying the function to $\sem{t}$.
While it is natural to express the desugarings of qualifiers with meta-level lambdas, our current framework does not allow meta-level abstractions. Instead, we extended the core language with macros: \cor{\id{MVar}}, \cor{\id{MLam}} and \cor{\id{MApp}}, that provide textual substitution. Core language macros allow us to approximate a higher-order meta-language.

The generic comprehension desugaring rules used in GHC actually allow for arbitrary monads, not just lists. Guards are desugared to sequential compositions $(>>)$ and a guard operation for the monad.  We omitted these constructs since for lists they are directly definable.  

The following table lists the intended desugaring rules, where $\%u$ is a freshly chosen name that is guaranteed to be unique and not used elsewhere in the term,.

\begin{minipage}{\linewidth}
\centering

\[\small\begin{array}{rcl}
          \sem{ \id{SListComp}(t,q) } &= & \mcor{\id{MApp}}(\sem{q},\sem{t})\\
          \sem{ \id{QEmpty} } &= & \mcor{\id{MLam}}( \%u, \mcor{\id{Return}}(\mcor{\id{MVar}}(\%u)))\\
          \sem{ \id{QBind}(\lit{x},t,q) } &=& \mcor{\id{MLam}}( \%u,  \mcor{\id{Bind}}(\sem{t},\mcor{\id{CLam}}([\lit{x}],\mcor{\id{MApp}}(\sem{q},\mcor{\id{MVar}}(\%u)))))\\
         % \sem{ \id{QGuard}(t,q) }&=& \mcor{\id{MLam}}( \%u,  \mcor{\id{Bind}}(\mcor{\id{Guard}}(\sem{t}),\mcor{\id{CLam}}([\%\_], \mcor{\id{MApp}}(\sem{q},\mcor{\id{MVar}}(\%u)))))\\
%          \sem{ \id{QGuard}(t,q) }&=& \mcor{\id{MLam}}( \%u,  \mcor{\id{Bind}}(\mcor{\id{CIf}}(\sem{t},\mcor{\id{CList}}([\mcor{\id{CBool}}(true)]),\mcor{\id{CList}}([])),\mcor{\id{CLam}}([\%\_], \mcor{\id{MApp}}(\sem{q},\mcor{\id{MVar}}(\%u)))))\\
% this can be even simpler
          \sem{ \id{QGuard}(t,q) }&=& \mcor{\id{MLam}}( \%u,  
          \mcor{\id{CIf}}(\sem{t}, 
          \mcor{\id{MApp}}(\sem{q},\mcor{\id{MVar}}(\%u)),
          \mcor{\id{CList}}([])))\\
          \sem{ \id{QLet}(\lit{x},t,q) } &=& \mcor{\id{MLam}}( \%u, \mcor{\id{Let}}(\lit{x},\sem{t},\mcor{\id{MApp}}(\sem{q},\mcor{\id{MVar}}(\%u))))
\end{array}\]
\end{minipage}

As we extended the sequential learning process of the previous case study, we assumed that the translation rules of the whole Pidgin language are fixed in advance. 

\textit{Hypothesis space.}
Our hypothesis space $\mathcal{H}^{lc}$ is mostly identical with $\mathcal{H}^1$, but it does not contain any of the meta-rules \mr{Case}, \mr{Throw}, \mr{Unzip} or \mr{Fresh}. Instead, to express the desugaring rules in this table, a macro version of the fresh rule was necessary:

\begin{displaymath}
\prftree[r]{\textsc{Meta-lambda}}{\Gamma ; i : \texttt{Id} \vdash M : \mcor{\sort{CTerm}}}{\Gamma  \vdash \mathbf{let}\, i = \mathtt{gensym}() \, \mathbf{in} \, \cor{\id{MLam}(\lit{i}, \lit{M})}:\mcor{\sort{CTerm}}}
\end{displaymath}

\textit{Decomposition.}
This case study demonstrates that sometimes "large" (>2) constructor groups are necessary, and our enumerative method does not scale up to large groups.

The smallest constructor group required for any example is $\src{SListComp}$ and $\src{QEmpty}$, so they need to be included in the first step. But any list comprehension example that does not use the other qualifiers simply reduces to $\cor{\id{Return}}$, therefore we can not learn the intended semantics of this group without the others. Adding the simplest qualifier, $\src{QGuard}$, is still not enough: without any bound variable we can not write examples that exclude similarly erroneous desugarings. Setting up the first step of the sequential learning task with just $\src{SListComp}$ and $\src{QEmpty}$ or perhaps adding $\src{QGuard}$ demonstrates a case where we would commit to a wrong translation rule early, and can not continue the learning process.

The first step needs to contain at least three source constructors, one of them should be $\src{QBind}$ or $\src{QLet}$. But this group is too large, with combined AST size of 14 our search runs to a timeout.

As a second test, we used a hint from the user: we provided the semantics (desugaring rule) of $\src{SListComp}$. This allowed us to learn the semantics of each qualifier one-by-one.

\textit{Results.}
The results we obtained after the user hint are presented in Table \ref{tab:list_comp_seq}.

\begin{table}[tb]
\small
\begin{tabular}{l | c | c | c | l}
Constructor group & AST size & Index & \#tests & time \\
\hline \hline
\src{QEmpty} & 3 & 6       & 1 & 0.0s \\
\src{QBind}  & 10 & 97632734 & 1 & 25min27s \\
\src{QLet}   & 7 & 116747 & 1 & 2s \\
\src{QGuard} & 6 & 31307 & 1 & 1s \\
total  &          && 4 & 25min30s 
\end{tabular}
\caption{Case study -- List comprehensions}
\label{tab:list_comp_seq}
\end{table}

In one case,  $\src{\id{QGuard}}$, the found semantics is equivalent with the intended one but shorter:
\[\small\begin{array}{rcl}
          \sem{ \id{QGuard}(t,q) }&=& \mcor{\id{CIf}}(\sem{t},\sem{q},\mcor{\id{MLam}}(\%\_,\mcor{\id{CList}}([])))\\
\end{array}\]

\textit{Test cases.}
We only needed one test case for every individual learning task: one test case suffices per source constructor. These test cases are not trivial: to demonstrate the non-trivial scoping behaviour of qualifiers we needed to embed them. To hypothetically auto-generate such test cases the binding and scoping rules of the source language must be taken into account.

\subsection{Try/Catch/Finally}

\textit{Setup.}
Finally we extended the Pidgin languages with exceptions. The source Pidgin is extended with try/catch/finally, which desugars to the core language extended with just try/catch.

\begin{minipage}{\linewidth}
\centering
\[\small
  \begin{array}{rcl}
    t, t_1 , t_2, t_3  \in \msrc{\sort{STerm}} &::= & \dots \bnfor
    \msrc{ \id{STryCatchFinally}(t_1,\lit{i},t_2,t_3) } \bnfor 
    \msrc{ \id{SThrow}(t) }\\
    e, e_1 , e_2, e_3  \in \mcor{\sort{CTerm}} &::= & \dots \bnfor
    \mcor{ \id{CTryCatch}(e_1,\lit{i}, e_2) } \bnfor 
    \mcor{ \id{CThrow}(e) }
\end{array}\]
\end{minipage}

In a $\msrc{\id{CTryCatchFinally}(t_1,\lit{i},t_2,t_3)}$ expression, 
first is  $t_1$ is executed.  
If it terminates normally with value $v$ then $t_3$ is executed with default outcome $v$.  If it raises an exception $ex$, then $t_2$ is executed with $\lit{i}$ bound to $ex$.  If executing $t_2$ terminates normally with value $v$ then $t_3$ is executed with default outcome $v$.  Otherwise if $t_2$ also raises an exception $ex'$ then $t_3$ is executed with default outcome $ex'$.  (In particular, the absence of an  exception handler can be emulated by having the handler be $\msrc{\id{Throw}(\lit{i})}$.)  When the \verb|finally| expression $t_3$ is executed, if it terminates normally with some result value, that value is ignored and the default outcome is performed instead.  If $t_3$ also raises an exception then this exception is the result and the default outcome is ignored.

Our intended desugaring is the following:

\begin{minipage}{\linewidth}
\centering

\[\small\begin{array}{rcl}
          \sem{ \id{STryCatchFinally}(t_1,\lit{i},t_2,t_3) } &= &
          \mcor{\id{CLet}}(\lit{\%v},\mcor{\id{CTryCatch}}(\mcor{\id{CTryCatch}}(\sem{t_1},\lit{i},\sem{t_2})),\\
&& \qquad         \lit{\%j},\mcor{\id{CLet}}(\%\_,\sem{t_3},\mcor{\id{CThrow}}(\mcor{\id{CVar}}(\lit{\%j})))),\\
          && \mcor{\id{CLet}}(\%\_,\sem{t_3},\mcor{\id{CVar}}(\lit{\%v})))\\
          \sem{ \id{CThrow}(t) } &=& \mcor{\id{CThrow}(\sem{t})}
\end{array}\]
\end{minipage}

Note that this desugaring rule duplicates the translated finally block $\sem{t_3}$. To avoid code bloat, one would normally create a thunk for the finally block. But our method can not distinguish the thunked version from inserting the finally block twice, and the latter is smaller. We use $\cor{\id{CLet}}$ with an unused variable for sequencing operations.

\textit{Hypothesis space.}
To express the intended desugaring, we need a version of the \mr{Fresh} rule for the new try-catch construct:

%\begin{displaymath}
%\prftree[r]{Fresh-let}{\Gamma \vdash M}{\Gamma ; \src{Var}(\lit{i}) : \texttt{STerm} \vdash N : \sigma}{\Gamma  \vdash \mathbf{let}\, i = \mathtt{gensym}() \, \mathbf{in} \, \cor{\id{CLet}(\lit{i},\lit{M},\lit{N}) }}
%\end{displaymath}\\
%\begin{displaymath}
%\prftree[r]{Fresh-trycatch}{\Gamma \vdash M}{\Gamma ; \src{Var}(\lit{i}) : \texttt{STerm} \vdash N : \sigma}{\Gamma  \vdash \mathbf{let}\, i = \mathtt{gensym}() \, \mathbf{in} \, \cor{\id{TryCatch}(\lit{M},\lit{i},\lit{N}) }}
%\end{displaymath}

\begin{displaymath}
\prftree[r]{\textsc{Fresh-trycatch}}{\Gamma \vdash M : \mcor{\sort{CTerm}}}{\Gamma ; x : \mcor{\sort{CTerm}} \vdash N : \mcor{\sort{CTerm}}}{\Gamma  \vdash \mathbf{let}\, i = \mathtt{gensym}() \, \mathbf{in} \, \cor{\id{CTryCatch}}(\lit{M},\lit{i},\lit{N}[\mcor{\id{Var}}(i)/x]) : \mcor{\sort{CTerm}}}
\end{displaymath}

Our first hypothesis space was based on $\mathcal{H}^1$, but we removed \mr{Case}, \mr{Throw}, and \mr{Unzip}, and added \mr{Fresh-trycatch}. But we found that the size of the intended desugaring is too large for such a general hypothesis space. 

We created a second, severely restricted hypothesis space $\mathcal{H}^{tcf}$. It only contained \mr{Fresh}, \mr{Fresh-trycatch}, and the core term constructors \cor{CThrow} and \cor{CTryCatch}. We also used an additional user layer: \mr{Fresh} and \mr{Fresh-trycatch} was in the first layer, so they were not allowed inside the core term constructors.

\textit{Decomposition.}
The relabelling of \src{SThrow} can easily be learned on its own, because throw has a unique behaviour. This allows us to learn in two steps.

\textit{Results.} 
The desugaring rule for \src{STryCatchFinally} (even without thunking) is very large, and we reached the timeout even with the very restricted hypothesis space $\mathcal{H}^{tcf}$. To see how much we missed the mark we ran the search to completion. The results are shown in Table \ref{tab:try_catch_seq}.

\begin{table}[tb]
\small
\begin{tabular}{l | c | c | c | l}
Constructor group & AST size & Index & \#tests & time \\
\hline \hline
\src{SThrow} & 2 & 18       & 1 & 0.0s \\
\src{STryCatchFinally}  & 13 & 396643849 & 7 & 1hours 16min 54s \\
total  &         & & 8 & 1hours 16min 54s
\end{tabular}
\caption{Case study -- Try-catch-finally}
\label{tab:try_catch_seq}
\end{table}

\textit{Test cases.} Learning the desugaring rule of \src{STryCatchFinally} needs many complex examples. First, try-catch-finally can branch: the main block may or may not throw and the catch block also may or may not throw. Covering all cases already needs 3 examples. We also need to ensure the evaluation order. In general, branching constructs needs at least as many examples as potential paths of execution. Also, to fix evaluation order we need examples that exclude any other permutation, and with constructs with many parameters ($\src{\id{STryCatchFinally}}$ has the most parameters amongst our constructs: 4) the number of potential permutations are higher.

\subsection{Threats to validity}\label{sec:threats}

Perhaps the most important, but at the same time the hardest thing to evaluate is how well our approach generalises. We can not yet evaluate our methods on real life programming languages%: we need to simplify them
, because real desugarings are too large. Pre-defined examples always carry the danger of inadvertent cherry-picking of the results. A further danger is that since we know the intended desugarings in advance, we do not need to experiment with user guidance such as restrictions on the hypothesis space, and we risk inadvertently underestimating the amount of user guidance needed.

We try to highlight each form of user guidance that we used in our case studies: each shows a limitation of our approach. We do not include writing the correct test cases or decomposing the language, since they were assumed to be the user's task from the beginning. 

\begin{enumerate}
    \item Our selection of meta-rules was somewhat tailored to the requirements. For example, we did not include a \mr{Fresh}-like rule for  \cor{CLam} because we did not need it.
    \item We needed a hand-written meta-rule \mr{Unzip} to express the desugaring rule of \src{SFor}. Other meta-rules used can be imagined to be auto-generated, but the \mr{Unzip} rule is hand-crafted.
    \item In two cases, \src{SBetween} and \src{STryCatchFinally}, we used a severely restricted hypothesis space, and in the second case we even used a user-defined layer. This shows the scalability problem: the desugaring rules have AST size 14 and 13 respectively, which is too large for our method to find with an unrestricted hypothesis space.
    \item Similarly, we needed user guidance to provide the rule of \src{SListComp}. This was again a scalability problem, although a different one: we would have to learn the rules for too many term constructors at once.
\end{enumerate}
We do not regard these limitations as fatal flaws.  Instead, they may illustrate that our approach should be thought of as a ``desugaring synthesis assistant'', rather than a fully automatic synthesis tool.  Knowledgeable semantics engineers may be able to make use of our approach's brute-force search even if detailed guidance is sometimes needed, analogously to interactive proof assistants in theorem proving.
In this light our work may be regarded as identifying an approach that may work, but further research (e.g. usability studies) would need to be done to determine whether our approach offers significant benefits in real semantics engineering efforts.

\section{Related work}

%The immediate context and basis of the present work was introduced by
%\citet{Krishnamurthi:2019}. They observed that many so-called \emph{tested semantics} can be split into a core language and a desugaring transformation, and proposed modelling such desugarings as tree transducers. They also presented four attempts to inductively learn such desugarings: a naive tree matching algorithm, a tree matching algorithm using Gibbs sampling, genetic programming, and program synthesis. The first two approaches relied on examples in the form of source programs translated to the core language, which is --- as they also highlighted --- not a reasonable requirement. The last two are more generic algorithms, but still were not able to handle cases such as \src{\id{SBetween}} and \src{\id{SFor}}.

%We picked up this thread and investigated the direction most promising to our judgement: their fourth attempt, program synthesis. Our  contributions are a formalisation which allows splitting the search space into feasible sub-tasks, desugaring extension problems, and an alternative definition of desugarings which highlights the role of the meta-language: the computational model in which we express translation rules. We showed how to use the sort systems of the languages to restrict the hypothesis space, and how to specify a set of translations with meta-program templates that can handle list transformations and  name generation.

Program synthesis is a vast research field, for a review see \citet{Gulwani:2017}. To solve the desugaring extension problem, we experimented with enumerative program synthesis, which may serve as a baseline algorithm, but we did not use extensive pruning or heuristics which may allow us to scale up learning to larger examples. In general, it is not clear how more sophisticated program synthesis methods could be applied to the desugaring extension problem. For example, in the desugaring extension problem there can be multiple unknown language constructs and they may appear in argument position in the examples. The implications of these differences need to be investigated. Most algorithms in program synthesis are specialised to a fixed language --- it is an open question how to abstract away the target language and what additional information on the target language is needed. Many algorithms are not specialised to inductive synthesis, and thus it is an open question whether any such method leads to a significant speed-up in our use-case.  

The most closely related work, by \citet{Krishnamurthi:2019}, has been discussed and compared with our approach throughout the paper.  There is existing work to automatically synthesise translations between languages, such as \emph{verified lifting}~\cite{Kamil:2016,Ahmad:2019}, but they search for translations of source \emph{programs} (of a fixed source language) to a fixed target language, as opposed to searching for translation rules for the source \emph{language}, which is a rather different problem.  Nevertheless there may be interesting connections between verified lifting and desugaring synthesis, which should be explored.

We are aware of a number of works that could serve as the basis of further investigation. FlashMeta~\cite{Polozov:2015} is an industrial framework by Microsoft to build  synthesis algorithms for user-defined target languages. The framework is based on inductive enumerative synthesis, thus it is possible that it could be used for our problem. The framework allows the user to define witness functions for the target language, that capture part of the inverse behaviour of the functions in the language and can speed up the search. A possible future direction is investigating whether providing suitable witness functions for a given core language is feasible, and whether it speeds up the search in our examples.

\citet{Bartha:2019} used \emph{meta-interpretive learning}~\cite{Muggleton:2014}, a framework for inductive logic programming, to learn the small-step semantic rules of a very simple programming language. The task they solved is close to ours; they attempted to learn inductively the structural operational semantics rules of a language rather than desugaring. However, their approach relied on identifying problem-specific \emph{meta-rules} (needed by meta-interpretive learning), so it is hard to see how their method can be generalised to various language features or larger languages.

In our code we represented the search space by sorted terms, a.k.a. algebraic data types (ADTs) in typed functional languages such as ML or Haskell. We are aware of two bodies of work that consider synthesising functions on ADTs: \emph{type-directed synthesis} as in Myth~\cite{osera} and Myth2~\cite{Frankle:2016}, and an extension of the Sketch framework~\cite{solar-lezama13ijsttt} called \emph{SyntRec}~\cite{Inala:2017}. Type-directed synthesis as implemented in Myth and Myth2 is not easy to apply to our problem because it requires the example set to be closed under recursive calls (for example, to learn a recursive function on a list we need examples showing the results of that function on all sub-lists).  On the other hand, SyntRec is more directly applicable to our problem, indeed it was the system used in the fourth attempt analysed by \KLE, but SyntRec was only successfully applied to languages without state. SyntRec, when applied to desugaring synthesis problems, also requires the core language interpreter to be implemented in the synthesis language, whereas our approach assumes only an opaque implementation of the core language is available.
Nevertheless, further evaluation is needed to see whether these systems can be modified to provide full solutions to the Pidgin challenge problem, or whether insights from their approaches can be incorporated into our approach.
%to compare whether for our use-case these methods are faster: their prominent examples has conceptual differences compared to ours.

Tested semantics such as $\lambda_{JS}$ for JavaScript~\cite{guha10ecoop}, S5~\cite{li15onward}, and $\lambda_{Py}$~\cite{Politz:2013} provide case studies showing how to handle large, real languages by desugaring to a core language.  Compared to Pidgin, these desugarings are considerably larger, and from inspecting their code or formalization artifacts, it is clear that some of the desugaring rules are much too large for our enumerative approach to handle (e.g. desugaring JavaScript function declarations to $\lambda_{JS}$ requires around 50 lines of Haskell code).  The Scheme report~\cite{r6rs} also specifies certain constructs by desugaring to a core, and may be a more plausible intermediate goal.  Interestingly, \citet{li15onward} identify the issue of \emph{semantic bloat}, resulting from desugarings that defensively cover all cases, but yield large amounts of boilerplate that is not needed in common cases.  This suggests an alternative strategy for learning complex desugarings, mirroring the gradual inductive approach apparently followed in practice, in which we might first seek simple explanations for common constructs in common cases (e.g. a ``Newtonian model''), and then look for counterexamples to the simple model and try to repair it to accommodate them (e.g. a ``relativistic model'').  Another interesting possibility is to analyze (non)interdefinability of core language features (see e.g. \cite{Spiders}) to guide the design of the core language or structure the decomposition into incremental learning problems.
We leave these intriguing possibilities for future work.

\section{Conclusion and future work}

Developing correct semantics rules for real-world languages is a necessary, but arduous, prerequisite to formal analysis. The problem of learning desugarings from examples has been introduced by \citet{Krishnamurthi:2019}, but the four approaches they tried each had inherent limitations. In this paper we have carefully analysed the problem and highlighted two key aspects: decomposability of language feature learning into a sequence of easier \emph{desugaring extension problems}, and careful attention to the hypothesis space and meta-language in which the desugaring rules are expressed. %Different desugaring extensions seem to require different meta-languages, and we show that decomposing desugaring learning into several independent extension problems and trying simpler meta-languages first leads to significant speedups compared to a brute-force approach of trying to learn multiple unrelated language features simultaneously.
Moreover, we show that, with some additional guidance, a simple enumerative search technique can successfully solve the desugaring synthesis challenge problem introduced by \citet{Krishnamurthi:2019}, and we evaluated our approach on additional extensions such as simple list comprehensions and exception handling with `finally'. While this is just one step toward the vision of learning the next 700 language semantics, our experimental results provide grounds for optimism. We plan to explore whether our approach can be incorporated into a CEGIS loop to automatically learn both semantics rules and suitable examples; whether the choice of suitable hypothesis spaces or meta-template rules can be automated; and whether our approach can be scaled up to synthesise desugarings for larger, more realistic languages that have been developed by hand, such as JavaScript~\cite{guha10ecoop} and Python~\cite{Politz:2013}.

\section*{Acknowledgements}

We would like to thank the anonymous reviewers and our shepherd Shriram Krishnamurthi for helpful feedback and suggestions for improvement. We also thank the artifact evaluators and AEC chairs for allowing us to update our artifact submission to reflect the final version of the paper.  This work was supported by ERC Consolidator Grant Skye (grant number 682315) and by an ISCF Metrology Fellowship grant provided by the UK government’s Department for Business, Energy and Industrial Strategy (BEIS).   Vaishak Belle was supported by a Royal Society University Research Fellowship.

\bibliography{bibfile}
\bibstyle{plainurl}
\bibliographystyle{plainnat}

%Uncomment to build tech report version
%\end{document}
\appendix

\newpage

\section{Reference implementation of Pidgin} \label{appendix:reference_impl}

For concreteness, we describe our reference implementation of the Pidgin source language, core language, the core language interpreter, and the intended desugaring translation in a Haskell-like pseudocode.  The source language behavior is defined as the composition of the desugaring and core language interpreter (though in our implementation we give a direct definition).  Our implementation (submitted as part of the supplementary material) shows the full details as runnable (but considerably more complex and less readable) Haskell code.  The differences are largely due to the requirements of using FEAT for enumerating terms.

\begin{figure}[h!]
\begin{verbatim}
newtype Id = Id String 

data Op = UMin | Not | Plus | Minus | And | Or | LT | GT 

data STerm = SVar String 
           | SPrim Op [STerm] 
           | SBetween STerm STerm STerm 
           | SNum Int 
           | SStr String
           | STrue 
           | SFalse 
           | SIf STerm STerm STerm
           | SLet Id STerm STerm
           | SLetRec Id STerm STerm
           | SLam [Id] STerm
           | SApp STerm [STerm]
           | SAssign [Id] STerm
           | SList [STerm]
           | SListCase STerm STerm STerm 
           | SFor STerm [(Id,STerm)] STerm

data CTerm = CVar Id 
           | CPrim1 Op CExp 
           | CPrim2 Op CExp
           | CNum Int
           | CStr String
           | CBool Bool
           | CIf CTerm CTerm CTerm
           | CLet Id CTerm CTerm
           | CLetRec Id CTerm CTerm
           | CLam [Id] CTerm
           | CApp CTerm [CTerm]
           | CAssign Id CTerm
           | CList [CTerm]
           | CListCase CTerm CTerm CTerm
           
rename :: Id -> Id -> CTerm -> CTerm
\end{verbatim}
\caption{Source and target term languages}\label{fig:syntax}
\end{figure}

\begin{figure}[p]
\begin{verbatim}
type Env = Map Id CTerm
type Result a = ... -- error and state monad
raise :: Error -> Result a
fresh :: Result Id
read :: Id -> Result CTerm
write :: Id -> CTerm -> Result ()
doUnop :: Op -> CTerm -> Result CTerm
doBinop :: Op -> CTerm -> CTerm -> Result CTerm
run : Result a -> Env -> Either a Error
\end{verbatim}
\caption{Helper functions for evaluation and desugaring}\label{fig:helper}

\begin{verbatim}
eval :: CTerm -> Either CTerm Error
eval t = run (eval' t) emptyEnv

eval' :: Map Id CTerm -> CTerm -> Result CTerm
eval' (CVar x)   = do
  e <- read x  -- might be an expression bound by letrec
  eval' e
eval' (CPrim1 op e) = do
  v <- eval' e
  case op of
    UMin -> doUnOp (negate @Int) v
    Not  -> doUnOp not v
    _    -> raise TypeError
eval' (CPrim2 op e1 e2) = do
  v1 <- eval' e1
  v2 <- eval' e2
  case op of
    Plus  -> 
      case v1 of
        (CNum _) -> doBinOp ((+) @Int) v1 v2                  
        (CList l1) -> case v2 of
                        CList l2 -> return (CList (l1 ++ l2))
                        _ -> raise TypeError
        _ -> raise TypeError
    Minus -> doBinOp ((-) @Int) v1 v2
    And   -> doBinOp (&&) v1 v2
    Or    -> doBinOp (||) v1 v2
    LT    -> doBinOp ((<) @Int) v1 v2
    GT    -> doBinOp ((>) @Int) v1 v2
    _ -> raise TypeError
eval' ex@(CNum _) = return ex
eval' ex@(CStr _) = return ex
eval' ex@(CBool _) = return ex
\end{verbatim}
\caption{Reference interpreter for \corelang, part 1}\label{fig:corelang-interp1}
\end{figure}

\begin{figure}[p]
\begin{verbatim}
eval' (CIf ex1 ex2 ex3) = do
  c <- eval' ex1
  case c of
    CBool b  -> if b then eval' ex2 else eval' ex3
    _ -> raise TypeError
eval' (CLet i ex1 ex2) = do
  v <- eval' ex1  -- let is eager
  i2 <- fresh
  write i2 v
  eval' (rename i id ex2)
eval' (CLetRec i ex1 ex2) = do
  i2 <- fresh
  write i2 (rename i id ex1)  -- any recursive references will be expanded later
  eval' (rename i id ex2)
eval' ex@(CLam _ _) = return ex
eval' (CApp ex es) = do
  f <- eval' ex
  case f of
    CLam is body -> do
        vs <- mapM eval' es
        case length is == length vs of
          True -> do iterM write (zip is vs)
                     eval' body
          False -> case vs of
                     [ CLang' (CList vs')] ->
                        case length is == length vs' of
                          True -> do iterM write (zip is vs')
                                     eval' body
                          False -> raise ArgumentNumberMismatchError
                     _ -> raise ArgumentNumberMismatchError
    _ -> raise TypeError
eval' (CAssign x ex) = do
  v <- eval' ex
  write x v
  return v
eval' (CList es) = do
  vs <- mapM eval' es
  return (CList vs)
eval' (CListCase e1 e2 e3) = do
  l <- eval' e1
  case l of
    CList []     -> eval' e2
    CList (e:es) -> do
          f <- eval' e3
          eval' (CApp f [e,(CList es)])
    _ -> raise TypeError
\end{verbatim}
\caption{Reference interpreter for \corelang, part 2}\label{fig:corelang-interp2}
\end{figure}

\begin{figure}[p]
\begin{verbatim}
ds :: STerm -> Either CTerm Error
ds t = run (ds' t) emptyEnv

ds' :: STerm -> Result STerm
ds' (SVar i) = return (CVar i)
ds' (SNum n) = return (CNum n)
ds' (SStr s) = return (CStr s)
ds' (SIf x y z) = do
     x' <- ds' x
     y' <- ds' y
     z' <- ds' z
     return (CIf x' y' z')
ds' (SLet i x y ) = do
     x' <- ds' x
     y' <- ds' y
     return (CLet i x' y')
ds' (SLetRec i x y ) = do
     x' <- ds' x
     y' <- ds' y
     return (CLetRec i x' y')
ds' (SLam ids x) = do
     x' <- ds' x
     return (CLam ids x')
ds' (SApp x xs)) = do
     x'  <- ds' x
     xs' <- mapM ds' xs
     return (CApp x' xs')
ds' (SAssign i x)) = do
     x' <- ds' x
     return (CAssign i x')
ds' (SList xs) = do
     xs' <- mapM ds' xs
     return (CList xs')
ds' (SListCase x y z) = do
     x' <- ds' x
     y' <- ds' y
     z' <- ds' z
     return (CListCase x' y' z')
\end{verbatim}
\caption{Desugaring reference implementation, part 1: straightforward cases.}\label{fig:desugaring1}
\end{figure}

\begin{figure}[p]
\begin{verbatim}
ds' STrue = return (CBool True)
ds' SFalse = return (CBool False)
ds' (SPrim o [s]) = do
     s'  <- ds' s
     CPrim1(o,s')
ds' (SPrim o [s1,s2]) = do
     s1'  <- ds' s1
     s2' <- ds' s2
     CPrim2(o,s1',s2')
ds' (SPrim (_,_)) = errorResult (syntaxError)
ds' (SBetween x y z) = do
     x' <- ds' x
     y' <- ds' y
     z' <- ds' z
     v1 <- freshVar
     v2 <- freshVar
     v2 <- freshVar
     return (CLet(v1,x',CLet(v2,y',CLet(v3,z',
                      CAnd (CPrim2(LT, CVar v1, CVar v2),
                                   CPrim2(LT, CVar v2, CVar v3))))))
ds' (SFor x bs y) = do
     x'  <- ds' x
     bs' <- mapM dsBind bs
     y'  <- ds' y
     (ids, zs) <- unzip bs'
     return (CApp (x', [CLam(ids, y'), CList(zs)]))

dsBind :: (Id,STerm) -> Result (Id,CTerm)     
dsBind (i,x) = do
        x' <- ds' x
        return (i,x')   
\end{verbatim}
      \caption{Desugaring reference implementation, part 2: nontrivial cases.}\label{fig:desugaring2}
\end{figure}

\end{document}